\documentclass[twocolumn,prX,nofootinbib]{revtex4-1}
\pdfoutput=1
\usepackage{amsmath,amssymb,amsfonts,eucal,mathrsfs,graphicx,tensor,color}
\usepackage{hyperref}
\DeclareMathAlphabet{\mathpzc}{OT1}{pzc}{m}{it}

\newcommand{\beq}{\begin{equation}}
\newcommand{\beqn}[1]{\begin{equation*}#1\end{equation*}}
\newcommand{\beql}[1]{\begin{equation}\label{#1}}
\newcommand{\eeq}{\end{equation}}
\newcommand{\bseq}{\begin{subequations}}
\newcommand{\bseql}[1]{\begin{subequations}\label{#1}}
\newcommand{\eseq}{\end{subequations}}
\newcommand{\eq}[1]{\eqref{#1}}

\newcommand{\lto}{\leftrightarrow}
\newcommand{\iim}{\Leftrightarrow}

\newcommand{\lf}{\left}
\newcommand{\rf}{\right}
\newcommand{\tql}[1]{{\textquotedblleft #1\textquotedblright}}

\newcommand{\ft}{\protect\footnote}

\newcommand{\rt}{\sqrt}
\newcommand{\fr}{\frac}
\newcommand{\tn}{\tensor}

\newcommand{\ord}{{\boldsymbol{\CMcal{O}}}}

\newcommand{\al}{\alpha}
\newcommand{\be}{\beta}
\newcommand{\de}{\delta}
\newcommand{\De}{\Delta}
\newcommand{\ep}{\epsilon}

\newcommand{\ka}{\kappa}
\newcommand{\la}{\lambda}
\newcommand{\La}{\Lambda}
\newcommand{\Sg}{\Sigma}
\newcommand{\om}{\omega}

\newcommand{\wg}{{\,\wedge\;}}
\newcommand{\com}[1]{\big[#1\big]}

\newcommand{\ed}{\mathrm{d}}

\newcommand{\Dl}{\nabla}
\newcommand{\LieD}{{\large\text{\pounds}}}

\newcommand{\dA}{\mathrm{d}A}
\newcommand{\dr}{\mathrm{d}r}
\newcommand{\ds}{\mathrm{d}s}

\newcommand{\dv}{\mathrm{d}v}
\newcommand{\dSg}{\mathrm{d}\Sigma}
\newcommand{\dOm}{\mathrm{d}\Omega}

\newcommand{\intdx}[1]{\int\mathrm{d}^{#1}x\,}

\newcommand{\hatchi}{\hat{\chi}}
\newcommand{\proj}{\mathbb{P}}
\newcommand{\vecDl}{\vec{\nabla}}
\newcommand{\sproj}{{\underline{\proj}}}
\newcommand{\uvec}[1]{{\underline{#1}}}
\newcommand{\uvecDl}{{\underline{\nabla}}}

\newcommand{\Gae}{G_{\ae}}
\newcommand{\GN}{G_\text{\sc n}}
\newcommand{\met}{\mathsf{g}}
\newcommand{\hatmet}{\hat{\mathsf{g}}}
\newcommand{\Rie}{\mathpzc{R}\,}
\newcommand{\Ric}{\mathpzc{R}\,}%ic
\newcommand{\EinG}{\mathpzc{G}}
\newcommand{\ac}{\CMcal{S}}
\newcommand{\acEH}{\ac_\text{\sc eh}}
\newcommand{\lag}{\mathscr{L}}
\newcommand{\emT}{\mathpzc{T}}
\newcommand{\aeT}{\emT^{\ae}}
\newcommand{\ADM}{{\text{\sc adm}}}
\newcommand{\khor}{\text{\sc kh}}
\newcommand{\uhor}{\text{\sc uh}}

\newcommand{\Sph}{\boldsymbol{\mathcal{B}}}

\begin{document}
\title{Mechanics of universal horizons}
\author{Per Berglund}\email{per.berglund@unh.edu}
\author{Jishnu Bhattacharyya}\email{jishnu.b@unh.edu}
\author{David Mattingly}\email{dyo7@unh.edu}
\affiliation{Department of Physics, University of New Hampshire, Durham, NH 03824, USA}

\begin{abstract}
Modified gravity models such as Ho\v{r}ava-Lifshitz gravity or Einstein-{\ae}ther theory violate local Lorentz invariance and therefore destroy the notion of a universal light cone. Despite this, in the infrared limit both models above possess static, spherically symmetric solutions with ``universal horizons'' - hypersurfaces that are causal boundaries between an interior region and asymptotic spatial infinity. In other words, there still exist black hole solutions. We construct a Smarr formula (the relationship between the total energy of the spacetime and the area of the horizon) for such a horizon in Einstein-{\ae}ther theory. We further show that a slightly modified first law of black hole mechanics still holds with the relevant area now a cross-section of the universal horizon. We construct new analytic solutions for certain Einstein-{\ae}ther Lagrangians and illustrate how our results work in these exact cases. Our results suggest that holography may be extended to these theories despite the very different causal structure as long as the universal horizon remains the unique causal boundary when matter fields are added.
\end{abstract}

%\makeatletter
%\let\old@fpheader\@fpheader
%\renewcommand{\@fpheader}{\old@fpheader\hfill UNH-12-02}
%\makeatother
\maketitle
%***********************************************************************************************************************************************************************************************************
\section{Introduction}
In the four decades since the seminal work of Bardeen, Carter, and Hawking~\cite{BCH} on the laws of black hole mechanics, a tremendous amount of effort has gone in to understanding horizon behavior. From the discovery of Hawking radiation~\cite{HawkingRad}, and the recognition that the four laws have a thermodynamic interpretation~\cite{Bekenstein:1973ur}, to holography~\cite{'tHooft, Susskind} and its concrete realization through the AdS/CFT correspondence~\cite{Maldacena, Gubser, Witten}, the physics of horizons has provided useful information about quantum gravity. Using horizon thermodynamics and some mild assumptions about the behavior of matter, one can reverse the logic and reconstruct general relativity as the thermodynamic limit of a more fundamental theory of gravity~\cite{J:EEqstate, Verlinde, Padmanabhan}. Integral to horizon dynamics is the first law of black hole mechanics, which for the simplest Schwarzschild case, and the most similar to what we are interested in, is just
	\beq
	\de M_{\ADM} = \fr{\ka_{\khor}\,\de A_{\khor}}{8\pi\GN}~.
	\eeq
Here $M_{\ADM}$ is the ADM mass of the spacetime and $\ka_{\khor}$ and $A_{\khor}$ are the surface gravity and cross-sectional area evaluated on the Killing horizon, respectively. Identifying $(8\pi\GN)^{-1}\ka_{\khor}$ as the temperature of the horizon and the area with the entropy allows one to make the analogy with the first law of thermodynamics, $\de E = T \de S$. 

In order to have an appropriate formulation of thermodynamics for the black hole itself, as well as the combined system of the black hole and the exterior environment, it is necessary for the horizon to have an inherent entropy. If there was no entropy associated with the horizon, one could simply toss objects into the black hole and reduce the total entropy of the black hole  and exterior system, thereby violating the second law of thermodynamics. If, however, the horizon has an associated entropy, then one can reformulate the usual second law into the generalized second law
	\beq
	\de \lf({S_\text{outside} + S_\text{horizon}}\rf) \geqq 0.
	\eeq
The generalized second law and the thought experiments behind it imply that \textit{any} causal boundary in a gravitational theory should have an entropy associated with it. In general relativity, the entropy can be shown to be proportional to the area of a slice of the Killing horizon. However, if one includes higher curvature terms the entropy is still a function of the metric and matter fields evaluated on a slice of the Killing horizon, though no longer proportional to the surface area alone~\cite{Jacobson:1993vj}. 

A much stronger departure from  general relativity comes when one considers models that allow vacuum solutions with non-zero tensor fields besides the metric. In these models Lorentz symmetry and the equivalence principle are in general broken. A simple example of such a model is that of Einstein-{\ae}ther theory~\cite{JM:ae:intro}, which introduces an {\ae}ther vector field $u^a$ and a dynamical constraint which forces $u^a$ to be a timelike unit vector everywhere. The introduction of the {\ae}ther vector preserves general covariance, but allows for novel effects such as matter fields travelling faster than the speed of light~\cite{Jacobson:2000gw} and new gravitational wave polarizations that travel at different speeds~\cite{JM:aewaves}. Given certain choices of the action for the {\ae}ther, the theory can be made phenomenologically viable~\cite{J:ae:constraints,J:ae:status}, have positive energy~\cite{GJ:+E}, and be ghost free~\cite{Eling:2004dk}. In addition, the {\ae}ther vector establishes a preferred frame and causality can be imposed in that frame~\cite{MattinglyReview} by requiring that all matter excitations propagate towards the future, even if the momentum vector of an excitation is spacelike. Since there is a preferred frame, Lorentz invariance does not hold, nor do the usual Lorentz invariance based arguments that a spacelike momentum vector in one frame immediately imply the existence of past directed momentum vectors. Thus, the propagation faster than the speed of light does {\it not} violate causality.

Even though there is a notion of causality, it seems at first glance as if there would be no causal boundaries equivalent to an event horizon in the Einstein-{\ae}ther theory -- by coupling the {\ae}ther vector $u^a$ to matter kinetic terms, the matter Lagrangian can be chosen to make matter perturbations about flat space propagate arbitrarily fast. This is  incorrect -- causally separated regions of spacetime can exist even in this case. In the static, spherically symmetric and asymptotically flat solutions of Einstein-{\ae}ther theory found by Eling and Jacobson~\cite{EJ:aebh} there exists such a region, and the boundary of this region has been dubbed the \tql{universal horizon}\ft{This point was first noted by Sergey Sibiryakov at Peyresq, 2010.}. Since the notion of a causal boundary and infinite speed modes is counterintuitive, we give a brief explanation of why they can occur here, postponing an in-depth discussion of universal horizons until later.

Consider a static, spherically symmetric spacetime, and cover it with Eddington-Finkelstein type coordinates such that the metric takes the form
	\beq
	\ds^2 = -e(r)\dv^2 + 2f(r)\dv\dr + r^2\dOm_2^2~,
	\eeq
and the time translation Killing vector is $\chi^a = \{1, 0, 0, 0\}$. Now let $\Sg_U$ denote a surface orthogonal to the {\ae}ther vector $u^a$, so that $U$ is the ``{\ae}ther time'' generated by $u^a$ that specifies each hypersurface in a foliation. At asymptotic spatial infinity $\chi^a$ and $u^a$ coincide, but as one moves in towards $r = 0$ each $\Sg_U$ hypersurface bends down to the infinite past in $v$, eventually asymptoting to a 3 dimensional spacelike hypersurface on which $(u\cdot\chi) = 0$, which implies that the Killing vector $\chi^a$ becomes tangent to $\Sg_U$. This hypersurface is the universal horizon. It is a causal boundary, as any signal must propagate to the future in $U$, which is necessarily towards decreasing $r$ at the universal horizon. The surface is regular, and in fact Barausse, Jacobson and Sotiriou~\cite{BJS:aebh} have numerically continued the solution for metric and {\ae}ther fields beyond the universal horizon.

Since such a causal boundary exists, it is natural to speculate that there must be an entropy associated with the universal horizon as well. In spherical symmetry, one does not need to worry about the zeroth law of black hole mechanics, as the symmetry enforces that all geometric quantities are constant over the universal horizon automatically. Hence one can immediately proceed to derive a Smarr formula and a corresponding first law. There are subtleties, however, as the boundary data of the theory at infinity naively contains two parameters. However, as noted in~\cite{EJ:aebh, BJS:aebh}, the boundary data can be reduced to one parameter, which is the total mass of the solution, by requiring that the solution is regular outside the universal horizon (see section \ref{sec:soln} for details). If one considers only regular solutions, we show that there is a Smarr relation between the total mass and geometric quantities evaluated on the universal horizon. In particular, the resulting Smarr relation contains a contribution from the extrinsic curvature of the $\Sg_U$ hypersurface in addition to the standard surface gravity term. We further show using a scaling argument that if one considers a transition between two regular solutions then there is also a first law that {\it may} admit a thermodynamic interpretation. With certain choices of the Lagrangian for Einstein-{\ae}ther theory, we construct new, exact solutions and use those as examples to gain insight into the thermodynamics of the first law. 

It is important to note that in previous works, black hole thermodynamics has failed when the limiting speed of matter fields is finite, but not necessarily the speed of light. In these cases one can construct perpetuum mobiles that violate the second law~\cite{Dubovsky:2006vk, Eling:2007qd, Jacobson:2008yc}. We will not consider any specific matter action in this paper, as our purpose is simply to determine whether a first law of mechanics holds for the universal horizon so that a thermodynamic interpretation \textit{might} exist. However it is certainly possible that a true thermodynamic interpretation could only hold if the universal horizon is the only causal boundary for all fields. This could, for instance, be done if the Lagrangian for any matter fields contained higher derivatives that made the local speed of excitations infinite as the energy increased. From the point of view of effective field theory this is natural as one expects all terms consistent with the symmetries of the problem to appear in any operator expansion. 

The paper is organized as follows. We first provide the background for Einstein-{\ae}ther theory in Sec.~\ref{sec:all-about-ae}. In Sec.~\ref{sec:Smarr-1L} we construct a Smarr formula and first law for spherically symmetric solutions. We then use two new exact, analytic solutions as examples in Sec.~\ref{sec:soln} to verify the Smarr formula and the first law, as well as discuss the regularity of these solutions. We conclude with some more speculative comments on how to proceed to establish the thermodynamic connection and the obstacles that still remain.
%**********************************************************************************************************************************************************************************************************
\section{The Einstein-{\ae}ther theory}\label{sec:all-about-ae}
Einstein-{\ae}ther theory was originally constructed~\cite{JM:ae:intro} as a mechanism for breaking local Lorentz symmetry yet retaining as many of the other positive characteristics of general relativity as possible. In particular it is the most general action involving the metric and a unit timelike vector $u^a$ that is two-derivative in fields and generally covariant. General covariance is maintained by enforcing the unit constraint on $u^a$ via a Lagrange multiplier. Following the presentation in  \cite{GJ:+E} the action of Einstein-{\ae}ther theory is a sum of the usual Einstein-Hilbert action $\acEH$ and the {\ae}ther action $\ac_{\ae}$~\cite{JM:ae:intro}
	\beql{ac:ae}
	\ac = \acEH + \ac_{\ae} = \fr{1}{16\pi\Gae}\intdx{4}\rt{-\met}\; \left(\Rie + \lag_{\ae}\right).
	\eeq
In terms of the tensor $\tn{Z}{^{ab}_{cd}}$ defined as\ft{Note the indicial symmetry $\tn{Z}{^{ba}_{dc}} = \tn{Z}{^{ab}_{cd}}$.}
	\beql{def:Zabcd}
	\tn{Z}{^{ab}_{cd}} = c_1\met^{ab}\met_{cd} + c_2\tn{\de}{^a_c}\tn{\de}{^b_d} + c_3\tn{\de}{^a_d}\tn{\de}{^b_c} - c_4u^au^b\met_{cd}~,
	\eeq
where $c_i,\,i = 1,\ldots,4$ are \tql{coupling constants} (or couplings, for short) of the theory, the {\ae}ther Lagrangian $\lag_{\ae}$ is given by
	\beql{lag:ae}
	-\lag_{\ae} = \tn{Z}{^{ab}_{cd}}(\Dl_a u^c)(\Dl_b u^d) - \la(u^2 + 1).
	\eeq
The {\ae}ther Lagrangian is therefore the sum of all possible terms for the {\ae}ther field $u^a$ up to mass dimension two, and a the constraint term $\la(u^2 + 1)$ with the Lagrange multiplier $\la$ implementing the normalization condition\ft{We use a convention where the metric has signature $(-, +, +, +)$.}
	\beql{ae:norm}
	u^2 = -1.
	\eeq
An additional term, $\Ric_{ab} u^a u^b$ is a combination of the above terms when integrated by parts, and is not included here.

There exists a number of theoretical as well as observational bounds on the couplings $c_i,\,i = 1,\ldots,4$ -- see e.g. \cite{J:ae:constraints, J:ae:status} for a comprehensive review. In this work, we  assume the following constraints to hold on these couplings
	\beql{c_i:constraints}
	0 \leqq c_{14} < 2, \qquad 2 + c_{13} + 3c_2 > 0, \qquad  c_{13} < 1~, 
	\eeq
where we have defined $c_{13} = (c_1 + c_3)$ and $c_{14} = (c_1 + c_4)$. As we will see, these combinations of couplings, as well as $c_{123} = (c_1 + c_2 + c_3)$, play a more direct role in our analysis than the individual couplings $c_i$.

The constraints \eq{c_i:constraints} come from the following conditions. If $c_{14} \geqq 2$ gravity becomes repulsive and one loses the proper Newtonian limit. Furthermore, in addition to the usual spin-2 gravitons, Einstein-{\ae}ther theory also possesses two vector and one scalar modes (corresponding to the three degrees of freedom of $u^a$). If $c_{14} < 0$ or $2 + c_{13} + 3c_2 < 0$ then the scalar mode squared speed (see \eq{spin-0:s0} below) about flat space becomes negative, signaling an instability of flat space to the production of scalar {\ae}ther-metric excitations. Also, $2 + c_{13} + 3c_2$ cannot be strictly zero, as the $G_\text{cosmo}$ appearing in the Friedmann equations derived from the Einstein-{\ae}ther theory needs to be positive and finite~\cite{Carroll:2004ai}. Similarly, if $c_{13} \geqq 1$ then the squared speed of the usual spin-2 graviton in flat space becomes infinite or negative, which generates the same problem but with the usual spin-2 graviton modes. As we will also see, the Smarr formula and the first law of black hole mechanics that we derive below becomes unphysical if $c_{13} = 1$. There are other observational limits on the couplings, e.g., coming from the requirement that propagating high energy cosmic rays do not lose energy due to vacuum \v{C}erenkov radiation of gravitons~\cite{Elliott:2005va}. We will explicitly not impose this constraint here as we are interested in the behavior of the scalar mode, the interplay of any scalar mode horizon with the Killing and universal horizons, and the possible role of \v{C}erenkov radiation from the universal horizon. Allowing the scalar mode to have any speed from almost zero to infinity is therefore theoretically useful.

The constant $\Gae$ in the action~\eq{ac:ae} is related to $\GN$, Newton's gravitational constant, via
	\beql{def:GN}
	\Gae = \lf(1 - \fr{c_{14}}{2}\rf)\GN~,
	\eeq
as can be established using the weak field/slow-motion limit of the Einstein-{\ae}ther theory~\cite{Carroll:2004ai}.

The equations of motion, obtained by varying the action \eq{ac:ae} with respect to the metric, $u^a$, and $\lambda$, are
	\beql{EOM}
	\EinG_{a b} = \aeT_{a b}, \qquad \AE_a = 0, \qquad u^2 = -1,
	\eeq
respectively, where the {\ae}ther stress tensor $\aeT_{ab}$ is given by
	\beql{ae:Tab}
	\begin{split}
	\aeT_{ab} = & \la u_a u_b + c_4 a_a a_b - \fr{1}{2}\met_{ab}\tn{Y}{^c_d}\Dl_c u^d + \Dl_c\tn{X}{^c_{a b}} \\
	& + c_1 \lf[(\Dl_a u_c)(\Dl_b u^c) - (\Dl^c u_a)(\Dl_c u_b)\rf]~,
	\end{split}
	\eeq
and to make the notation more compact we have introduced the following
	\beql{def:AE}
	\begin{split}
	& \AE_a = \Dl_b\tn{Y}{^b_a} + \la u_a + c_4(\Dl_a u^b)a_b~, \\
	& \tn{Y}{^a_b} = \tn{Z}{^{ac}_{bd}}\Dl_c u^d~, \\
	& \tn{X}{^c_{a b}} = \tn{Y}{^c_{(a}}\tn{u}{^{}_{b)}} - \tn{u}{^{}_{(a}}\tn{Y}{_{b)}^c} + u^c Y_{(ab)}~.
	\end{split}
	\eeq
The acceleration vector $a^a$ appearing in the expression for the {\ae}ther stress tensor is defined as the parallel transport of the {\ae}ther field along the {\ae}ther field\ft{We use the conventional notation $\Dl_X$ for the directional derivative ($X^a\Dl_a$) along any vector field $X^a$. Once the normalization condition \eq{ae:norm} is imposed, the acceleration is always orthogonal to the {\ae}ther field (i.e. $u\cdot a = 0$), and therefore, is always spacelike.}
	\beql{def:a}
	a^a = \Dl_u u^a~.
	\eeq
%===========================================================================================================================================================================================================
\subsection{Static, spherically symmetric expansions}
We now turn to the static, spherically symmetric case and provide some useful expansions which we will later use to analyze the equations of motion. Eling and Jacobson observed that for any spherically symmetric solution of Einstein-{\ae}ther theory, $u^a$ is hypersurface orthogonal\ft{In fact they are also solutions of Ho\v{r}ava gravity~\cite{Horava, J:hso-ae=hor}.}~\cite{EJ:aebh}, which in turn implies that the twist of $u^a$ vanishes, i.e.
	\beql{HSO:0}
	u_{[a}\Dl_b u_{c]} = 0.
	\eeq
An immmediate consequence of the hypersurface orthogonality condition is that there exists a one-parameter redundancy among the couplings $c_i$. Using the unit norm constraint \eq{ae:norm}, the squared twist can be expressed as
	\beql{twist^2}
	\om^2 = (\Dl_a u_b)(\Dl^a u^b) - (\Dl_a u_b)(\Dl^b u^a) + a^2~,
	\eeq
which also vanishes for hypersurface orthogonal solutions. We can, therefore, add any multiple of $\om^2$ to the action without affecting the solutions. In particular, by adding
	\beql{ac:twist^2-term}
	\De\ac = -\fr{c_0}{16\pi\Gae}\intdx{4}\rt{-\met}\;\om^2
	\eeq
to the action \eq{ac:ae}, where $c_0$ is an arbitrary real constant, the sole effect would be to obtain a new {\ae}ther Lagrangian $\lag'_{\ae}$, otherwise identical to \eq{lag:ae}, except with a new set of coupling constants $c'_i$, $i = 1, ..., 4$, related to the un-primed $c_i$ through
	\beql{c-redef}
	c'_1 = c_1 + c_0, \;\;\; c'_2 = c_2, \;\;\; c'_3 = c_3 - c_0, \;\;\; c'_4 = c_4 - c_0.
	\eeq
Thus, by appropriately choosing $c_0$, one can set any one of the couplings  $c_1$, $c_3$ and $c_4$, or any appropriate combinations of them, to any preassigned value. On the other hand, the coupling $c_2$, as well as combinations like $c_{13}$, $c_{14}$ and $c_{123}$ \eq{c_i:constraints} stay invariant under the above  redefinition of the couplings \eq{c-redef}. 

Our analysis will be further facilitated by defining a set of basis vectors at every point in spacetime so that we can project out various components of the equations of motion. Staticity and spherical symmetry implies the existence of a time translation Killing vector $\chi^a$ as well as three rotational Killing vectors $\zeta_{(i)}^a, i = 1, 2, 3$ (only two of the three are linearly independent). It is often convenient to choose $\chi^a$ as one of the basis vectors, but in this case it is actually more helpful to use a different basis. We first take $u^a$ to be the (timelike) basis vector. We next pick any two spacelike unit vectors, call them $m^a$ and $n^a$, both of which are normalized to unity, are mutually orthogonal and lie on the tangent plane of the two-spheres $\Sph$ that foliate the hypersurface $\Sg_U$. Finally, we use $s^a$, the spacelike unit vector orthogonal to $u^a$, $m^a$ and $n^a$ that points ``outwards'' along a $\Sg_U$ hypersurface. Note that the acceleration $a^a$ only has a component along $s^a$ by spherical symmetry, i.e.
	\beql{reln:a-s}
	a^a = (a\cdot s) s^a~.
	\eeq
Thus, our tetrad consists of $\{u^a, s^a, m^a, n^a\}$. By spherical symmetry, any physical vector may have components along $u^a$ and $s^a$, while any rank-two tensor may have components along the bi-vectors $u_a u_b$, $u_{(a} s_{b)}$, $u_{[a} s_{b]}$, $s_a s_b$ and $\hatmet_{a b}$, where $\hatmet_{a b}$ is the projection tensor onto the two-sphere $\Sph$, bounding a section of a $\Sg_U$ hypersurface. For example, the basis-expansion of the extrinsic curvature of a $\Sg_U$ hypersurface is
	\beql{basis-expand:Kab}
	K_{a b} = K_0 s_a s_b + \fr{\hat{K}}{2} \hatmet_{a b}~,
	\eeq
where $K_0$ and $\hat{K}$\ft{As a consequence of spherical symmetry and $u^a$ being orthogonal on the hypersurface $\Sg_U$, we have $s \wg \ds = 0$ -- compare with \eq{HSO:0}. Thus the vectors $s^a$ are orthogonal to the hypersurfaces $\{\Sg_s\}$, foliating the spacetime. It can then be shown that $\hat{K}$ is the trace of the extrinsic curvature of the two-spheres $\Sph$ due to their embedding in $\Sg_s$. See appendix \ref{append} for further details.} are scalar parameters related to each other through
	\beql{reln:K-K0-hatK}
	K = K_0 + \hat{K}~,
	\eeq
with $K$ the trace of the extrinsic curvature of the $\Sg_U$ hypersurface. We ask the reader to refer to appendix \ref{append} for further details on these points.

Finally, we work out some geometric quantities related to the Killing vector $\chi^a$. We first use the spherical symmetry to write the timelike Killing vector $\chi^a$ as
	\beql{basis-expand:chi}
	\chi^a = -(u\cdot\chi)u^a + (s\cdot\chi)s^a~.
	\eeq
The basis-expansion of $\Dl_a\chi_b$ takes the following form
	\beql{basis-expand:Dlchi}
	\Dl_a \chi_b = -\ka(u_a s_b - s_a u_b)~.
	\eeq
Here $\ka$ is defined as the surface gravity on any two-sphere $\Sph$ since, as we will now show, (at least outside the Killing horizon) $\ka$ is the acceleration of a static observer on $\Sph$ as measured by an observer at asymptotic infinity. First consider the region outside the Killing horizon where there exists an outward pointing spacelike unit vector $r^a$, the radial unit vector, orthogonal to $\chi^a$. The tangent vector $\hatchi^a$ to the world line of a static observer on any two-sphere $\Sph$ is simply the unit timelike vector along $\chi^a$. In terms of the redshift factor $\rho$, we can then write $\chi^a = \rho\hatchi^a$ (i.e. $\rho^2 = -\chi\cdot\chi$). Using \eq{basis-expand:Dlchi}, the directional derivative of $\hatchi^a$ along itself is given by $\Dl_{\hatchi}\hatchi^a = a_\chi r^a$, where $a_\chi = (\ka/\rho)$ is the local acceleration of the static observer. In other words, $\ka = \rho a_\chi$ is the redshifted acceleration with respect to an observer at infinity, which prompts us to call $\ka$ the surface gravity on the two-sphere $\Sph$. The mathematical analysis also follows through inside the Killing horizon\ft{Note however that inside the Killing horizon $\hatchi^a$ is spacelike and $r^a$ is timelike. Associating a spacelike unit vector with an observer is allowed here since there is no local limiting speed.} and the local acceleration is still given by $(\ka/\rho)$, but the interpretation of $\rho$ as the redshift factor no longer holds. Nevertheless, we will continue to call $\ka$ the surface gravity. Using the results from appendix \ref{append}, the surface gravity is explicitly given by\ft{We thank Ted Jacobson for pointing out reference~\cite{Jacobson-Parentani} in this context, where the surface gravity at the Killing horizon is generally shown to be proportional to the expansion of a congruence of timelike geodesics. However, we emphasize that in the present paper, equation~\eq{def:ka} holds everywhere in spacetime rather than just at the Killing horizon. We also note that using $(u\cdot\chi)_{\khor} = -(s\cdot\chi)_{\khor}$ at the Killing horizon, we have $\ka_{\khor} = -(u\cdot\chi)_{\khor}\{(a\cdot s) + K_0\}_{\khor}$ from \eq{def:ka}. However, this relation and the the central result of~\cite{Jacobson-Parentani}, although similar in appearance, actually differ since the {\ae}ther does not define a geodesic flow.}
	\beql{def:ka}
	\ka = \rt{-\fr{1}{2}(\Dl_a \chi_b)(\Dl^a \chi^b)} = -(a\cdot s)(u\cdot\chi) + K_0(s\cdot\chi)~.
	\eeq
	
As mentioned in the introduction, the universal horizon occurs when the Killing vector becomes tangent to a $\Sg_U$ hypersurface. Therefore as one travels inwards from spatial infinity along a $\Sg_U$ hypersurface, the universal horizon is actually reached only as a limit. Hence our quantities defined as ``on the universal horizon'' refer to this limit, rather than some actual intersection of $\Sg_U$ and the universal horizon which is a hypersurface of constant $r$ (again in Eddington-Finkelstein coordinates). On the universal horizon, $(u\cdot\chi)_{\uhor} = 0$ and $(s\cdot\chi)_{\uhor} = |\chi|_{\uhor}$ where $|\chi|_{\uhor}$ is the magnitude of the Killing vector $\chi^a$ on the universal horizon. Therefore the surface gravity on the universal horizon is
	\beql{def:ka:UH}
	\ka_{\uhor} = K_{0, \uhor}|\chi|_{\uhor}~.
	\eeq
By spherical symmetry, the surface gravity is constant over any two-sphere, and thus on the universal horizon as well. 
%===========================================================================================================================================================================================================
\subsection{Equations of motion for the static, spherically symmetric case}
We next  study   Einstein's equations and the {\ae}ther equations of motion \eq{EOM} by explicitly using the time translational and spherical symmetries of the problem, in addition to hypersurface orthogonality. To set up the Einstein's equations, we need to know the basis-expansion of the {\ae}ther stress tensor and the Ricci tensor. By spherical symmetry, they take the following form
	\beql{ae:Tab:K}
	\aeT_{a b} = \aeT_{u u}u_a u_b - 2\aeT_{u s}u_{(a} s_{b)} + \aeT_{s s}s_a s_b + \fr{\hat{\aeT}}{2}\hatmet_{ab}~,
	\eeq
and
	\beql{basis-expand:Rab}
	\Ric_{a b} = \Ric_{u u}u_a u_b - 2\Ric_{u s}u_{(a} s_{b)} + \Ric_{s s}s_a s_b + \fr{\hat{\Ric}}{2}\hatmet_{a b}~,
	\eeq
respectively. The coefficients of $\aeT_{a b}$ in \eq{ae:Tab:K} are computed from the general expression \eq{ae:Tab} for the stress tensor, using the results in appendix \ref{append}. The corresponding coefficients for $\Ric_{a b}$, on the other hand, are computed from the defining equation $\com{\Dl_a, \Dl_b}X^c = -\tn{\Rie}{_{a b d}^c}X^d$ by choosing $X^a = u^a$ or $s^a$, and then contracting the expression again with $u^a$ and/or $s^a$ appropriately\ft{We note that the coefficient $\hat{\Ric}$ cannot be constructed in this method. This is not an obstruction for most part since we do not need the explicit expression for $\hat{\Ric}$. In section \ref{sec:soln} we use the explicit coordinates to construct $\hat{\Ric}$.}. For our present purpose, it is sufficient to show the components $\aeT_{u s}$ and $\Ric_{u s}$, which are given as follows
	\beql{aeT-Ric:comps:us}
	\begin{split}
	& \aeT_{u s} = c_{14}(\hat{K}(a\cdot s) + \Dl_u (a\cdot s))~, \\
	& \Ric_{u s} = (K_0 - \hat{K}/2)\hat{k} - \Dl_s \hat{K}~,
	\end{split}
	\eeq
where $\hat{k}$ is the extrinsic curvature of the two-spheres $\Sph$ due to their embedding in $\Sg_U$ (see appendix \ref{append})\ft{Intimately related to this is the fact that $\hat{k}/2$ is the coefficient of $\hatmet_{a b}$ in the basis-expansion of $\Dl_a s_b$.}. Comparing \eq{ae:Tab:K} and \eq{basis-expand:Rab}, we see that there are altogether four\ft{We have one extra Einstein's equation, because we do not assume staticity here. Thus these equations can also be used to study time dependent but spherically symmetric perturbations around static solutions.} non-trivial components of the Einstein's equations. 

The {\ae}ther's equations of motion \eq{EOM}, on the other hand, reduce to the scalar equation\ft{After solving for the Lagrange multiplier there is no non-trivial projection of $\AE^a$ \eq{def:AE} along $u^a$.}
	\beql{ae:EOM:s-proj}
	\begin{split}
	0 = (s\cdot\AE) = c_{13}\Dl_s K_0 & + c_{13}(K_0 - \hat{K}/2)\hat{k} \\
	& + c_2\Dl_s K - \aeT_{us}~,
	\end{split}
	\eeq
as a consequence of hypersurface orthogonality and spherical symmetry. Quite naturally, the coupling constants that appear in \eq{ae:EOM:s-proj} above are precisely those which are invariant under \eq{c-redef}. 
 
A well-known fact in general relativity is that the Bianchi identities (a consequence of general covariance) can be used to show that a subset of the Einstein equations are actually constraint equations. In Einstein-{\ae}ther theory, which is also generally covariant, there are generalized Bianchi identities, and projections of these give rise to constraint equations as well. As explained in \cite{J:Ceqn} (see also~\cite{BJS:aebh}), the generalized Bianchi identities for Einstein-{\ae}ther theory are
	\beql{ae:EOM-Bianchi}
	\Dl^a\lf[\EinG_{a b} - \aeT_{a b} + u_a \AE_b\rf] + \AE_a\Dl_b u^a = 0~,
	\eeq
and the corresponding constraint equations for the $\Sg_U$ hypersurfaces are~\cite{BJS:aebh, J:Ceqn}
	\beql{ae:Ceqn:gen}
	\lf(\EinG_{a b} - \aeT_{a b}\rf)u^a  - \AE_b = 0~.
	\eeq
Projecting  \eq{ae:Ceqn:gen} along $s^b$ and using \eq{aeT-Ric:comps:us} and \eq{ae:EOM:s-proj} the explicit form of the constraint equation, adapted to the foliation $\Sg_U$, is
	\beql{ae:Ceqn}
	\begin{split}
	0 = c_{123}\Dl_s K_0 & - (1 - c_{13})(K_0 - \hat{K}/2)\hat{k} \\
	& + (1 + c_2)\Dl_s \hat{K}~.
	\end{split}
	\eeq

On the other hand, subtracting the equation $c_{13}(\Ric_{u s} - \aeT_{us}) = 0$\ft{This equation, obviously, follows from the Einstein equation $(\Ric_{u s} - \aeT_{us}) = 0$.} from the {\ae}ther's equation of motion \eq{ae:EOM:s-proj}, we get
	\beql{ae:scalar:EOM}
	c_{123}\Dl_s K = (1 - c_{13})\aeT_{us}~.
	\eeq
Equations \eq{ae:Ceqn} and  \eq{ae:scalar:EOM} are two independent linear combinations of the {\ae}ther's equation of motion \eq{ae:EOM:s-proj} and the Einstein equation $\Ric_{u s} = \aeT_{us}$. Therefore the two sets of equations are equivalent. In section \ref{sec:soln}, we study these equations along with the $uu$, $ss$ and the spherical components of the Einstein's equations.

Finally, considering the projection of the {\ae}ther equations of motion along the Killing vector $\chi^a$, we arrive at the following equation of central importance in this paper (as it will be used to derive a Smarr formula)
	\beql{Smarr:EOM}
	\Dl_b \CMcal{F}^{a b} = 0, \qquad \CMcal{F}_{a b} = q(u_a s_b - s_a u_b)~,
	\eeq
where the quantity $q$ is given by
	\beql{Smarr:def:q}
	\begin{split}
	q = -\lf(1 - \fr{c_{14}}{2}\rf)(a\cdot s)(u\cdot\chi) & + (1 - c_{13})K_0(s\cdot\chi) \\
	& + \fr{c_{123}}{2}K(s\cdot\chi)~.
	\end{split}
	\eeq
The derivation of \eq{Smarr:EOM} closely follows the manipulations leading to the Smarr formula in \cite{BCH}. We also found the algebraic relations of appendix \ref{append}, especially those discussed in the last part of the appendix, useful in arriving at \eq{Smarr:EOM}.

The similarity between \eq{Smarr:EOM} and the source free Maxwell's equations allows us to solve \eq{Smarr:EOM} exactly once we adopt a particular coordinate system. A useful choice is Eddington-Finklestein like coordinates (see equation \eq{met:static-sph-sym} in section \ref{sec:soln}), as these coordinates are good everywhere in the spacetime. Because of staticity and spherical symmetry, in this particular coordinate system $\CMcal{F}_{a b}$ has only one non-trivial component, namely $\CMcal{F}_{vr}$. Therefore, solving \eq{Smarr:EOM} amounts to solving for the electrostatic field of a point charge at $r = 0$ in this particular geometry. By Gauss's law, we   conclude $q \sim r^{-2}$. Using the asymptotic series solutions \eq{asymptotic-behaviour:all} we then fix the constant of proportionality and obtain
	\beql{Smarr:q:explicit}
	q = \lf(1 - \fr{c_{14}}{2}\rf)\fr{r_0}{2r^2}~,
	\eeq
where $r_0$ is {\it the} single free parameter which defines the regular Einstein-{\ae}ther black hole solutions. We ask the reader to refer to section \ref{sec:soln} for a more detailed discussion of how the parameter counting works and related issues. The parameter $r_0$, as we show in the following section, defines the mass of the Einstein-{\ae}ther black holes. With the aid of equations \eq{Smarr:EOM}, \eq{Smarr:def:q} and \eq{Smarr:q:explicit} and the fact that the Einstein-{\ae}ther black holes constitute a one-parameter family of solutions, one can provide very simple derivations of the Smarr relation and the first law for Einstein-{\ae}ther black hole mechanics, which we now turn to.
%**********************************************************************************************************************************************************************************************************
\section{The Smarr formula and  first law}\label{sec:Smarr-1L}
In this section we give very simple derivations of the Smarr formula and the first law of (universal) horizon mechanics for a general static and spherically symmetric Einstein-{\ae}ther black hole.

To begin with, we need a suitable definition of the mass of the Einstein-{\ae}ther black holes. In this case, the ADM mass of a solution is identical to its Komar mass. From the general definition of the Komar mass of a stationary solution
	\beql{def:MADM}
	M_{\ADM} = -\fr{1}{4\pi\Gae}\int\limits_{\Sph_\infty}\dSg_{ab}\Dl^a \chi^b~,
	\eeq
where $\dSg_{ab}$ is the integration measure on any two-sphere $\Sph$, explicitly given by
	\beqn{
	\dSg_{ab} = -u_{[a} s_{b]}\dA~,
	}
with $\dA$ the differential area element on the two-sphere $\Sph$, and $\Sph_\infty$ is the sphere at infinity. Note the apperance of $\Gae$ (as opposed to $\GN$) in \eq{def:MADM} -- this ensures the correct weak field/slow-motion limit of the Einstein-{\ae}ther theory, as we will see below.

We can further express the right hand side of \eq{def:MADM} in terms of the surface gravity on the sphere at infinity following \eq{basis-expand:Dlchi}. Using the asymptotic expressions of \eq{asymptotic-behaviour:all} -- in which a particular choice of coordinates have been made -- in \eq{def:ka}, we find
	\beql{MADM:aebh}
	\begin{split}
	M_{\ADM} & = -\fr{1}{4\pi\Gae}\int\limits_{\Sph_\infty}\dA\,(a\cdot s)(u\cdot\chi) \\
	& = \fr{1}{4\pi\Gae}\int\limits_{\Sph_\infty}\dA\,(a\cdot s) = \fr{r_0}{2\Gae}~.
	\end{split}
	\eeq
	
Our starting point of the derivation of the Smarr relation is equation \eq{Smarr:EOM}. As already noted, structurally \eq{Smarr:EOM} resembles the source free Maxwell's equations with $\CMcal{F}_{a b}$ akin to a purely electrostatic field. In particular, by Gauss' law the flux of $\CMcal{F}_{a b}$ through the sphere at asymptotic infinity equals the flux through sphere at the universal horizon\ft{According to \eq{append:Fab:flux}, the flux of $\CMcal{F}_{a b}$ through any two-sphere, $\Sph_r$, at radius $r$, is the surface integral of $q$ over $\Sph_r$. By spherical symmetry, this flux is the value of $q$ at $r$, i.e., $q(r)$, times the area of the sphere $\Sph_r$ itself.}. Performing the flux integrals and using the expression for the ADM mass \eq{MADM:aebh}, we arrive at the promised Smarr relation
	\beql{Smarr:MADM}
	\lf(1 - \fr{c_{14}}{2}\rf)M_{\ADM} = \fr{q_{\uhor} A_{\uhor}}{4\pi\Gae}~,
	\eeq
where $A_{\uhor}$ is the area of the universal horizon, $q_{\uhor}$ is the value of $q$ \eq{Smarr:def:q} on the universal horizon,
	\beql{def:quh}
	q_{\uhor} = (1 - c_{13})\ka_{\uhor} + \fr{c_{123}}{2}K_{\uhor}|\chi|_{\uhor}~,
	\eeq
and we have used the expression \eq{def:ka:UH} for the surface gravity at the universal horizon, $\ka_{\uhor}$, to write $q_{\uhor}$ as above. Following~\cite{Eling:aeE, Foster:aeNQ} and \cite{GJ:+E} we can furthermore introduce $M_{\ae}$, given by
	\beql{def:Mae}
	M_{\ae} = \lf(1 - \fr{c_{14}}{2}\rf)M_{\ADM}~,
	\eeq
which is the total mass of an asymptotically flat solution defined in the asymptotic {\ae}ther rest frame. Using \eq{MADM:aebh} and the relation \eq{def:GN} between $\GN$ and $\Gae$, we then have~\cite{Eling:aeE, Foster:aeNQ, GJ:+E}
	\beql{reln:Mae-r0-GN}
	M_{\ae} = \fr{r_0}{2\GN} \qquad\iim\qquad M_{\ae}\GN = M_{\ADM}\Gae~.
	\eeq
The above relation between $M_{\ae}$ and $r_0$ ensures that one gets the correct Newtonian limit of the Einstein-{\ae}ther theory far away from the sources~\cite{Carroll:2004ai, Eling:aeE, Foster:aeNQ}. In terms of the total mass, one also obtains a more natural presentation of the Smarr formula~\eq{Smarr:MADM}, namely
	\beql{Smarr}
	M_{\ae} = \fr{q_{\uhor} A_{\uhor}}{4\pi\Gae}.
	\eeq

To obtain the first law for the Einstein-{\ae}ther black holes, we need to consider a variation which takes us from a given regular solution to a distinct nearby regular solution. The key observation leading to the first law is that the regular Einstein-{\ae}ther black hole solutions depend on the single dimensionful parameter $r_0$ introduced in \eq{Smarr:q:explicit}. As a result, the location of the universal horizon, $r_{\uhor}$, is related to $r_0$ through a relation of the form $r_\uhor = \mu r_0$, where $\mu$ is a dimensionless quantity which can depend only on the coefficients $c_2$, $c_{13}$ and $c_{14}$. From the proportionality between $r_{\uhor}$ and $r_0$, we now have $q_{\uhor} \sim r_0^{-1}$ from \eq{Smarr:q:explicit}, while $A_{\uhor} = 4\pi r_{\uhor}^2 \sim r_0^2$, and hence $\de q_{\uhor}A_{\uhor} = -\fr{1}{2}q_{\uhor}\de A_{\uhor}$. Considering then a variation of the Smarr relation \eq{Smarr}, the first law for Einstein-{\ae}ther black holes follows in a straightforward manner~\cite{Townsend}
	\beql{1L}
	\de M_{\ae} = \fr{q_{\uhor} \de A_{\uhor}}{8\pi\Gae}~.
	\eeq

Note that our derivation of the Smarr formula and the first law makes it manifest that at least when spherical symmetry is present, we can always have a \tql{first law} applied to any sphere\ft{This point has also been stressed in~\cite{EJ:aebh}, where a first law for the {\ae}ther black holes applied to the {\it spin-0 horizon} was obtained. For earlier work on the first law for an {\ae}ther black hole at the Killing horizon using the Noether approach~\cite{Wald:NQ}, see~\cite{Foster:aeNQ}.}, where a variation of the total mass is proportional to $q$ evaluated on the sphere, times the variation of the area of the sphere. Furthermore, since on dimensional grounds $q \sim r_0^{-1}$ as well as $\ka \sim r_0^{-1}$, where $q$ and $\ka$ are the values of the respective quantities evaluated on the sphere in question, we can always write such a first law in terms of the surface gravity on the sphere. However, the importance of the universal horizon rests on the fact that it is {\it the} causal boundary in the spacetime and therefore, only \eq{1L} should possibly have a thermodynamic interpretation.
%**********************************************************************************************************************************************************************************************************
\section{The solutions}\label{sec:soln}
In this section, our main goal is to present two exact, asymptotically flat, static, spherically symmetric, single parameter families of {\ae}ther black hole solutions. These solutions provide additional evidence for the general result~\cite{EJ:aebh, BJS:aebh} that all asymptotically flat, static and  spherically symmetric {\ae}ther black holes depend on a single parameter  after imposing a regularity condition (to be discussed below). This single-parameter dependence, as already emphasized earlier, is crucial in our derivation of the first law. With our exact solutions, we furthermore verify the Smarr formula \eq{Smarr} and the first law \eq{1L}. As we will also see, the interesting piece in $q_{\uhor}$ \eq{def:quh}, that depends on the trace of the extrinsic curvature at the universal horizon, is absent (for separate reasons) for both these special solutions.

In the following, we first adopt a convenient coordinate system to express the equations of motion (see the paragraph following \eq{ae:scalar:EOM}). Next, we present an asymptotic series solution of these equations, valid for large $r$ and for arbitrary nonzero values of the couplings $c_2$, $c_{13}$, $c_{123}$ and $c_{14}$. The asymptotic solution has already been obtained in previus work~\cite{EJ:aebh, BJS:aebh}, and our purpose of presenting it here is three-fold: First of all, the asymptotic analysis determines the asymptotic (and sometimes the exact) nature of various relevant functions in the problem (e.g. the functional form $q$ in \eq{Smarr:q:explicit}) and allows us to obtain the ADM mass \eq{MADM:aebh}. Secondly, the asymptotic analysis reveals that the general {\ae}ther black hole solution can depend on at most two parameters, thereby providing a natural route to the topic of regularity of the solutions. Finally, the asymptotic analysis for two special choices of coupling constants lead to the exact solutions mentioned above. We discuss these special solutions in subsections \ref{subsec:exact-c14} and \ref{subsec:exact-c123}, respectively.

To perform an asymptotic analysis of the equations, we  set up an Eddington-Finklestein-like coordinate system which naturally respects the symmetries of the problem. With this choice of coordinates, the metric takes the form
	\beql{met:static-sph-sym}
	\ds^2 = -e(r)\dv^2 + 2f(r)\dv\dr + r^2\dOm_2^2~,
	\eeq
and the timelike Killing vector is
	\beql{chi:static-sph-sym}
	\chi^a = \{1, 0, 0, 0\}~.
	\eeq
The {\ae}ther field can be parametrized as
	\beql{u:static-sph-sym}
	u^a = \lf\{\al(r), \be(r), 0, 0\rf\}, \quad \be(r) = \fr{e(r)\al(r)^2 - 1}{2f(r)\al(r)}~,
	\eeq
where the relation between $\al(r)$ and $\be(r)$ takes care of the unit norm constraint \eq{ae:norm}. Therefore, to perform the asymptotic analysis, we only need the asymptotic behaviour of the three functions $e(r)$, $f(r)$ and $\al(r)$, which are given as follows
	\beql{asymptotic-behaviour:e-f-u}
	\begin{split}
	e(r) = 1 + & \ord(r^{-1}), \quad f(r) = 1 + \ord(r^{-1})~, \\
	& \al(r) = 1 + \ord(r^{-1})~.
	\end{split}
	\eeq
The boundary conditions on the metric coefficients are such that the solution is asymptotically flat, while those for the {\ae}ther components are such that (the boundary condition on $\al(r)$ implies $\be(r) = \ord(r^{-1})$ asymptotically)
	\beql{asymptotic-behaviour:aether}
	\lim_{r \to \infty} u^a = \{1, 0, 0, 0\}~.
	\eeq

It is worthwhile to make the following comments at this stage: we set up the asymptotic analysis for arbitrary non-zero values of the coefficients $c_2$, $c_{13}$ and $c_{14}$ such that $c_{123} \neq 0$. Although the results we present do not show this explicitly, if one takes the limit $c_{123} \to 0$ of this solution, most of the coefficients in the $1/r$ expansions of $e(r)$, $f(r)$ and $\al(r)$ diverge. Therefore, we need to set up a separate asymptotic analysis when $c_{123} = 0$ (but $c_{14} \neq 0$), and we then discover the exact solution presented in section \ref{subsec:exact-c123}. On the other hand, if one takes the $c_{14} \to 0$ of the general asymptotic solution by keeping $c_{123}$ non-zero, the exact solution presented in section \ref{subsec:exact-c14} is obtained\ft{The scalar equation of motion \eq{ae:scalar:EOM} becomes trivially satisfied if both $c_{123}$ and $c_{14}$ vanish simultaneously, and consequently the structure and solution space of the field equations changes significantly.}.

We can now solve the Einstein's equations and the {\ae}ther's equation of motion order by order in $1/r$. We have solved the relevant equations to $\ord(r^{-10})$ which determines  the functions $e(r)$, $f(r)$ and $\al(r)$ to $\ord(r^{-8})$\ft{All the equations are second order ODEs in $r$ and hence at $\ord(r^{-(n + 2)})$ the functions only up to $\ord(r^{-n})$ contribute.}. Not surprisingly, the asymptotic forms of these functions, as well as those which depend on them are quite cumbersome and do not convey too much information beyond that they can be found. We thus quote the relevant results only up to $\ord(r^{-2})$. To begin with, the metric components are
	\bseql{asymptotic-behaviour:all}
	\beql{asymptotic-behaviour:all:e(r)}
	\begin{split}
	& e(r) = 1 - \fr{r_0}{r} + \ord(r^{-3})~, \\
	& f(r) = 1 + \fr{c_{14} r_0^2}{16 r^2} + \ord(r^{-3})~,
	\end{split}
	\eeq
while the components of the {\ae}ther are
	\beql{asymptotic-behaviour:all:al_1(r)}
	\begin{split}
	& \al(r) = 1 + \fr{r_0}{2 r} + \fr{3r_0^2 - 8r_{\ae}^2}{8r^2} + \ord(r^{-3})~, \\
	& \be(r) = -\fr{r_{\ae}^2}{r^2} + \ord(r^{-3})~.
	\end{split}
	\eeq
Our results agree perfectly with those in~\cite{BJS:aebh} (see equations 24$-$26), under the identification $F_1 \lto -r_0$ and $A_2 \lto (\fr{3}{8}r_0^2 - r_{\ae}^2)$.

Given the results in \eq{asymptotic-behaviour:all:e(r)} and \eq{asymptotic-behaviour:all:al_1(r)}, we can compute the series expansion for everything else; for example the asymptotic forms of $(a\cdot s)$, $(u\cdot\chi)$ and $(s\cdot\chi)$ are
	\beql{asymptotic-behaviour:all:a}	
	\begin{split}
	& (a\cdot s) = \fr{r_0}{2 r^2} + \ord(r^{-3})~, \\
	& (u\cdot\chi) = - 1 + \fr{r_0}{2 r} + \fr{r_0^2}{8 r^2} + \ord(r^{-3})~, \\
	& (s\cdot\chi) = \fr{r_{\ae}^2}{r^2} + \ord(r^{-3})~,
	\end{split}
	\eeq
respectively. The various components of the extrinsic curvature $K_{a b}$, as well as its trace, are likewise
	\beql{asymptotic-behaviour:all:K-K0-hatK}	
	\begin{split}
	& K_0 = \fr{2r_{\ae}^2}{r^3} + \ord(r^{-5})~, \\
	& \hat{K} = - \fr{2r_{\ae}^2}{r^3} + \ord(r^{-5}), \\
	& K = \ord(r^{-5})~.
	\end{split}
	\eeq
	\eseq
These results are useful to compute the ADM mass \eq{MADM:aebh}.

The solution, at this stage, depends on two parameters namely $r_0$ and $r_{\ae}$. Among these, the length scale $r_0$ is akin to the \tql{Schwarzschild radius} and is related to the total mass of the black hole according to \eq{reln:Mae-r0-GN}. The parameter $r_{\ae}$ is essentially the $\ord(r^{-2})$ coefficient of $\al(r)$, and is defined in this way for convenience. From the asymptotic analysis, it appears that $r_{\ae}$ is a second free parameter on which the solutions  depend. This, as we now explain following~\cite{EJ:aebh, BJS:aebh}, is not the case after all; rather $r_{\ae}$ is related to $r_0$ upon requiring that the solutions are regular everywhere outside the universal horizon.

We begin our discussion with the observation that the Einstein-{\ae}ther theory admits a \tql{ground state} solution where the spacetime is four dimensional Minkowski and the {\ae}ther is $\{1, 0, 0, 0\}$ with respect to an observer in the preferred frame (called the {\ae}ther rest frame). In \cite{JM:aewaves}, the authors consider perturbations around this background and show, in particular, that there is a spin-0\ft{The results of \cite{JM:aewaves} show that when the {\ae}ther is not hypersurface orthogonal and if no symmetry is assumed, there are additional spin-1 and spin-2 modes. In our case, the condition of hypersurface orthogonality of the {\ae}ther will prevent any spin-1 mode from propagating in the black hole backgrounds we consider. Likewise, any spin-2 mode will be excluded because of spherical symmetry.} mode which propogates with a speed $s_0$ given by
	\beql{spin-0:s0}
	s_0^2 = \fr{c_{123}(2 - c_{14})}{c_{14}(1 - c_{13})(2 + c_{13} + 3c_2)}~,
	\eeq
with respect to the {\ae}ther rest frame. Because of  general covariance of the Einstein-{\ae}ther theory, perturbations around an {\ae}ther black hole background will also give rise to a spin-0 mode with a local speed given by \eq{spin-0:s0}. The spin-0 horizon is a hypersurface beyond which any outward moving excitation travelling with $s_0$ (or less) gets trapped. More precisely, the spin-0 horizon is hypersurface where the timelike Killing vector becomes null with respect to the \tql{effective spin-0 metric} $\met^{(0)}_{a b} = \met_{a b} - (s_0^2 - 1)u_a u_b$~\cite{EJ:aebh, BJS:aebh}. Equivalently, we can also define the spin-0 horizon as the hypersurface where $(s\cdot\chi)^2 = s_0^2(u\cdot\chi)^2$.

For generic values of the couplings $c_2$, $c_{13}$ and $c_{14}$, which respect \eq{c_i:constraints}, the spin-0 speed $s_0$ is a non-zero finite quantity. Consequently, the spin-0 horizon can be located anywhere outside the universal horizon. However, for finite non-zero $s_0$ one can always apply the field redefinitions introduced in \cite{Foster:field-redef} to set $s_0 = 1$, thereby making the spin-0 horizon coincide with the Killing horizon~\cite{EJ:aebh, BJS:aebh}. This extra condition therefore reduces the number of independent couplings from three to two. However, we should emphasize that this does not mean we are exploring a restricted coupling space. Rather, we are using an extra freedom in the theory (the field redefinitions) to conveniently choose (by imposing $s_0 = 1$) typical sets of couplings $\{c_2, c_{13}, c_{14}\}$ which label larger equivalent classes. In \cite{EJ:aebh, BJS:aebh}, the authors use the above logic to effectively scan a smaller coupling space in their numerical constructions of asymptotically flat, static and spherically symmetric {\ae}ther black holes. Those studies clearly prove that a solution for generic values of $r_0$ and $r_{\ae}$ is singular, precisely at the location of the spin-0 horizon. However, once the solution is required to be regular everywhere outside the universal horizon, the extra constraint automatically makes $r_{\ae}$ dependent on $r_0$, i.e., the former cannot be an extra parameter. Since the general asymptotically flat solution can at most depend on two parameters, the regularity condition reduce the parameter space so that we have a one-parameter family of solutions. In this manner, \cite{EJ:aebh, BJS:aebh} obtain a unique asymptotically flat {\ae}ther black hole solution for a given value of the parameter $r_0$ (and a given set of couplings) by making the solution regular at the corresponding spin-0 horizon.

The field redefinitions of \cite{Foster:field-redef} (see also \cite{EJ:aebh}) mentioned above also scale $c_{123}$ and $(1 - c_{13})$ in the same way while keep $c_{14}$ invariant. It is then clear that such a transformation does not exist when either of $c_{123}$ or $c_{14}$ vanish or when $c_{13} = 1$ (we can rule out this last possibility owing to \eq{c_i:constraints}). At the same time, according to \eq{spin-0:s0}, the spin-0 speed diverges as $c_{14} \to 0$, while it vanishes as $c_{123} \to 0$\ft{We note that there are other  limits of \eq{spin-0:s0} when $s_0$ can vanish or diverge: $c_{14} \to 2$ ($s_0$ vanishes), $c_{13} \to 1$ ($s_0$ diverges) and $(2 + c_{13} + 3c_2) \to 0$ ($s_0$ diverges). However, they all violate the constraints \eq{c_i:constraints}, and therefore are excluded on physical grounds.}. In the context of a black hole solution, when $c_{14} = 0$, the spin-0 horizon coincides with the universal horizon, since as noted in the introduction, the latter is {\it the} causal boundary for arbitrarily fast excitations. On the other hand, when $c_{123} = 0$, the spin-0 horizon in a black hole solution is pushed all the way to spatial infinity and so overlaps with the asymptotic boundary. Owing to the absence of the field redefinitions for these cases however, the spin-0 horizons {\it cannot} be mapped on to the metric horizon. Remarkably, there exists exact solutions for these special cases, where the spin-0 regularity condition can be illustrated in an explicit manner. We present these solutions in the following two subsections.
%==========================================================================================================================================================================================================
\subsection{Exact solution for $c_{14} = 0$}\label{subsec:exact-c14}
When the coupling $c_{14}$ is set to zero, the system admits an exact solution given by
	\bseql{exact-soln:c14-0:e-f-u}
	\beql{exact-soln:c14-0:e-f}
	e(r) = 1 - \fr{r_0}{r} - \fr{c_{13}r_{\ae}^4}{r^4}, \qquad f(r) = 1~,
	\eeq
and
	\beql{exact-soln:c14-0:u}
	\begin{split}
	& \al(r) = \fr{1}{e(r)}\lf(-\fr{r_{\ae}^2}{r^2} + \rt{e(r) + \fr{r_{\ae}^4}{r^4}}\rf)~, \\
	& \be(r) = -\fr{r_{\ae}^2}{r^2}~.
	\end{split}
	\eeq
From the explicit solution, we  further work out
	\beql{exact-soln:c14-0:u.X-s.X}
	\begin{split}
	& (s\cdot\chi) = -\be(r) = \fr{r_{\ae}^2}{r^2}~, \\
	& (u\cdot\chi) = -\rt{e(r) + \be(r)^2} = -\rt{1 - \fr{r_0}{r} + \fr{(1 - c_{13})r_{\ae}^4}{r^4}}~.
	\end{split}
	\eeq
	\eseq
As mentioned in the beginning \eq{c_i:constraints}, we  always assume $c_{13} < 1$.

We  now investigate  the locations of the Killing and universal horizons in this solution. By definition, the Killing horizon is located where $(\chi\cdot\chi) = 0$, or equivalently, at the largest root of $e(r) = 0$, and the universal horizon is located at the largest root of $(u\cdot\chi) = 0$. From \eq{exact-soln:c14-0:e-f} and \eq{exact-soln:c14-0:u.X-s.X} this  amounts to solving two quartic equations. Rather than doing this directly, we extract most of the important properties of these roots from simple arguments.

We begin by noting that $e'(r) > 0$ everywhere owing to $c_{13} > 0$. Therefore, $e(r)$ is a monotonically increasing function and we have a single real root\ft{Of course, $e(r) = 0$ must have at least two real roots, but the second real root must be negative by the above argument, and hence unphysical.} at $r = r_\khor$, which is the location of the Killing horizon. Secondly, from \eq{exact-soln:c14-0:u.X-s.X} $(u\cdot\chi)^2 = e(r) + (s\cdot\chi)^2$ -- therefore from the monotonicity of $e(r)$, even the largest root of $(u\cdot\chi)^2$ is necessarily located at some $r = r_\uhor < r_\khor$. We also conclude that $(u\cdot\chi)^2 = 0$  has at most two real roots by noting that the function has a single minimum at $r = \rt[3]{4(1 - c_{13})r_{\ae}^4/r_0}$. Furthermore, the two roots are distinct when $r_{\ae} < r^*_{\ae}$, they coincide when $r_{\ae} = r^*_{\ae}$, and there are no real roots when $r_{\ae} > r^*_{\ae}$, where
	\beqn{
	r^*_{\ae} = \fr{r_0}{4}\lf[\fr{27}{1 - c_{13}}\rf]^{1/4}~.
	}
The situation should be contrasted with the existence of an event horizon for the usual charged (Reissner-Nordstrom) black hole. However, there is an important difference between the present solution and the charged black hole solution in regards with the regularity of the solutions everywhere. In case of a charged black hole, the solutions are regular  everywhere except at $r = 0$ and are physically allowed as long as the extremality condition is met. To examine the regularity of the present solution, we can see from the expressions of the curvature scalars for the present solution
	\beqn{
	\Rie = \fr{6c_{13}r_{\ae}^4}{r^6}, \qquad \Ric_{a b}\Ric^{a b} = \fr{90c_{13}^2r_{\ae}^8}{r^{12}}~,
	}
that the ambient spacetime is free of any curvature singularities except at $r = 0$. But, the {\ae}ther field being a physical component of the theory, we also need to make sure that the solution for the {\ae}ther is regular everywhere as well. A coordinate independent quantity associated with the {\ae}ther, which can signal the existence of pathologies in the present solution is $(u\cdot\chi)$\ft{$(s\cdot\chi)$ for this solution is another coordinate independent quantity. But it is nicely behaved everwhere except at $r = 0$, and is therefore incapable of signaling irregularities.}. Indeed, when $r < r_\uhor$, $(u\cdot\chi)^2$ is negative based on our discussion above, and hence $(u\cdot\chi)$ is purely imaginary between  $r = r_\uhor$ (the location of the universal horizon) and the smaller root of $(u\cdot\chi)^2 = 0$. In other words, for generic values of $r_{\ae}$, the {\ae}ther solution is irregular at the universal horizon. Naturally, this irregularity is prevented if $(u\cdot\chi)^2$ is never allowed to be negative. This regularity condition, along with the demand for the existence of at least one root\ft{As required by cosmic censorship -- since we have superluminal propagation, the Killing horizon cannot save cosmic censorship.} of $(u\cdot\chi)^2 = 0$, uniquely implies that the regular physical solution exists, iff
	\beql{def:rae}
	r_{\ae} = r^*_{\ae} = \fr{r_0}{4}\lf[\fr{27}{1 - c_{13}}\rf]^{1/4}~.
	\eeq
Thus, $r_{\ae}$ is not an independent parameter. We have already argued that the spin-0 horizon for this solution overlaps with the universal horizon. Therefore, the regularity condition is indeed a regularity condition at the spin-0 horizon.

	\begin{figure}[t!]
	\centering
	\includegraphics[scale=0.8]{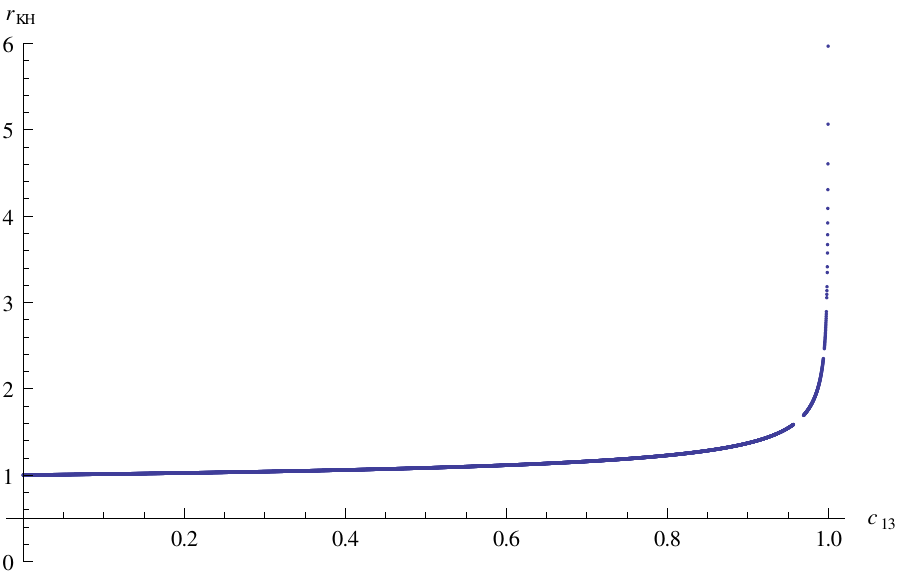}
	\caption{$r_\khor$ in units of $r_0$ as a function of $c_{13}$.}
	\label{figure:rkh-c13}
	\end{figure}  

Manifestly, the regular solution depends on a single parameter $r_0$. Here onwards, when we talk about this exact solution as well as about any quantity pertaining to it, the condition \eq{def:rae} will always be implied. The location of the universal horizon for this physical solution is very easy to find -- it is a root of both $(u\cdot\chi)^2 = 0$ and $\ed/\dr(u\cdot\chi)^2 = 0$, and is given by
	\beql{exact-soln:c14-0:r_uhor}
	r_{\uhor} = \fr{3r_0}{4}~.
	\eeq
Quite interestingly, the result does not depend on the value of $c_{13}$. The location of the Killing horizon $r_\khor$ however does depend on $c_{13}$, but we did not attempt to find an analytical expression for it; instead, we solved for $r_\khor$ numerically and the result is presented in figure~\ref{figure:rkh-c13}.

From its definition \eq{aeT-Ric:comps:us} $\aeT_{u s}$ vanishes when $c_{14} = 0$ and the {\ae}ther stress tensor becomes diagonal, with the nontrivial components given by
	\beq
	\aeT_{u u} = -\fr{3c_{13}r_{\ae}^4}{r^6}, \quad \aeT_{s s} = -\aeT_{u u}, \quad \hat{\aeT} = 4\aeT_{u u}~.
	\eeq
From the equation of motion \eq{aeT-Ric:comps:us} we then have $\Dl_sK= 0$, i.e., $K$ is constant on a given hypersurface $\Sg_U$. But $K$ vanishes asymptotically for asymptotically flat spacetimes -- this can be most readily seen from the fact that $u^a \sim \chi^a$ asymptotically, so that $K \sim \Dl\cdot\chi = 0$. Thus $K = 0$ on every hypersurface $\Sg_U$, and therefore everywhere in spacetime. A related point is that, manifestly, the solution does not depend on the coupling $c_2$ in any way. According to \cite{J:hso-ae=hor}, $c_2$ is the coupling of the $K^2$ term of the {\ae}ther Lagrangian. Therefore we see that every reference to $c_2$ has been removed as $K$ vanishes. 

We are now in a position to derive a Smarr relation. Given the solution \eq{exact-soln:c14-0:e-f-u} we can compute the surface gravity on the universal horizon following \eq{def:ka:UH} and this turns out to be
	\beql{exact-soln:c14-0:ka:UH}
	\ka_{\uhor} = \fr{8}{9r_0(1 - c_{13})}~.
	\eeq
Using $A_{\uhor} = 4\pi r_\uhor^2 = 9\pi r_0^2/4$ as the area of the universal horizon, we can therefore derive a  Smarr formula for the present solution\ft{Note: $M_{\ae} = M_{\ADM}$ and $G_{\ae} = \GN$ for this solution since $c_{14} = 0$.}
	\beql{exact-soln:c14-0:Smarr}
	M_{\ae} = \fr{(1 - c_{13})\ka_{\uhor}A_{\uhor}}{4\pi\Gae}~.
	\eeq
Varying the parameter $r_0$ we then also obtain a first law of black mechanics for the present solution explicitly
	\beql{exact-soln:c14-0:1L}
	\de M_{\ae} = \fr{(1 - c_{13})\ka_{\uhor}\,\de A_{\uhor}}{8\pi\Gae}~.
	\eeq
Comparing \eq{exact-soln:c14-0:Smarr} and \eq{exact-soln:c14-0:1L} with the general Smarr relation \eq{Smarr} and the first law \eq{1L} respectively, we find perfect agreement. Naturally, the interesting piece proportional to $c_{123}K_\uhor$ is absent as $K = 0$ everywhere in this solution.
%==========================================================================================================================================================================================================
\subsection{Exact solution for $c_{123} = 0$}\label{subsec:exact-c123}
When the coupling $c_{123}$ is set to zero, we also have an exact solution given by
	\beql{exact-soln:c123-0:e-f-u}
	\begin{split}
	& e(r) = 1 - \fr{r_0}{r} - \fr{r_u(r_0 + r_u)}{r^2}, \qquad f(r) = 1~, \\
	& \al(r) = \lf(1 + \fr{r_u}{r}\rf)^{-1}, \qquad \be(r) = -\fr{r_0 + 2r_u}{2r}~,
	\end{split}
	\eeq
where $r_u$ is a {\it positive} constant, given by
	\beql{def:ru}
	r_u = \lf[\rt{\fr{2 - c_{14}}{2(1 - c_{13})}} - 1\rf]\fr{r_0}{2}~.
	\eeq
The requirement of positivity of $r_u$ follows from demanding the function $\al(r)$ be regular everywhere. Consequently, this imposes the following bound on the couplings $c_{13}$ and $c_{14}$
	\beql{exact-soln:c123-0:c_13-c_14}
	c_{14} \leqq 2 c_{13}~.
	\eeq
Therefore, for this special case we also need to ensure that $c_{13}$ is non-negative in addition to $ c_{13} < 1$ and $0 \leqq c_{14} < 2$ as we assume in general \eq{c_i:constraints}. The positivity of $r_u$ criterion also rules out another possible solution for $r_u$
	\beqn{
	r_u = -\lf[\rt{\fr{2 - c_{14}}{2(1 - c_{13})}} + 1\rf]\fr{r_0}{2}~,
	}
which solves all the equations of motion, but is manifestly negative for all values of $c_{13}$ and $c_{14}$. The curvature invariants for this solution are
	\beqn{
	\Rie = 0, \qquad \Ric_{a b}\Ric^{a b} = \fr{4r_u^2(r_0 + r_u)^2}{r^8}~,
	}
demonstrating that the geometry is free from any curvature singularity everywhere except at $r = 0$. Thus the solution is regular everywhere.

The reader might have spotted that the parameter $r_{\ae}$ does not appear in this solution. Instead we have the parameter $r_u$, which is however not a free parameter. Rather, there are two possible choices for $r_u$ as functions $r_0$ and this ambiguity is resolved (in the form of choosing $r_u$ positive) by demanding that the solution be regular everwhere except at $r = 0$. One way to appreciate the difference between the present solution and all the solutions with $c_{123} \neq 0$, is to note that we need separate asymptotic analysis for the cases $c_{123} \neq 0$ and $c_{123} = 0$\ft{There is no such need when the coupling in question is $c_{14}$ as the general asymptotic solution \eq{asymptotic-behaviour:all} admits a smooth $c_{14} \to 0$ limit.}. As it turns out, when $c_{123} \neq 0$, the equations of motion force the $\ord(r^{-1})$ coefficient of $\al(r)$ to be $r_0/2$ \eq{asymptotic-behaviour:all:al_1(r)}. When $c_{123} = 0$, this requirement no longer holds and we are left with a free parameter $r_u$. Furthermore, when $c_{123} \neq 0$, the parameter $r_{\ae}$ appears as the $\ord(r^{-2})$ coefficient of $\al(r)$ and stays as the second free parameter in the general asymptotic analysis. For the $c_{123} = 0$ case however, the $\ord(r^{-2})$ coefficient is a function of $r_u$, and as the subsequent analysis reveals, the equations of motion eventually restrict $r_u$ to take one of the two choices mentioned above. The requirement for regularity then removes one of the choices. We have already argued earlier that the spin-0 horizon for this solution \tql{coincides} with the asymptotic boundary at infinity. We thus see that even with the correct boundary conditions imposed, there can be two different solutions corresponding to two different values of the paremeter $r_u$\ft{The choice of the correct asymptotic boundary conditions with appropriate fall off conditions for the various fields should be treated as partly making the spin-0 horizon regular in this case. In particular, this seems to be the reason for no additional {\it continuous} parameter in the solution.}. The actual regularity condition here comes in the form of choosing the correct value of $r_u$ as given in \eq{def:ru}. Manifestly, the regular solution depends on a single parameter $r_0$. 

We next adress the issue of the locations of the Killing and universal horizons. From the explicit solution, we  further work out
	\beql{exact-soln:c123-0:u.X-s.X}
	(s\cdot\chi) = -\be(r) = \fr{r_0 + 2r_u}{2r}, \qquad (u\cdot\chi) = -1 + \fr{r_0}{2r}~.
	\eeq
The roots of $e(r) = 0$ can be explicitly found in this case and they are $r = (r_0 + r_u)$ and $r = -r_u$ respectively. The second root is negative, i.e. unphysical, and the unique Killing horizon is at
	\beql{exact-soln:c123-0:rkh}
	r_{\khor} = r_0 + r_u~.
	\eeq
On the other hand, $(u\cdot\chi)$ has a unique root at $r = r_0/2$ which is therefore the location of the universal horizon in this solution
	\beql{exact-soln:c123-0:r_uhor}
	r_{\uhor} = \fr{r_0}{2}~.
	\eeq
As in the case of the exact solution with $c_{14} = 0$, the location of the universal horizon does not depend on the coupling constants for the present case either, while the location of the Killing horizon does.

With $\aeT_{us}$ vanishing according to \eq{ae:scalar:EOM} the stress tensor is also diagonal in this solution, with the non-trivial components given by
	\beq
	\begin{split}
	& \aeT_{u u} = -\fr{2c_{13}(r_0 + 2r_u)^2 - c_{14}r_0^2}{8r^4}, \\
	& \aeT_{s s} = -\aeT_{uu}, \qquad \hat{\aeT} = 2\aeT_{uu}~.
	\end{split}
	\eeq

To derive a Smarr relation, we proceed as before and compute the surface gravity on the universal horizon from \eq{def:ka:UH} using the present solution \eq{exact-soln:c123-0:e-f-u}
	\beql{exact-soln:c123-0:ka:UH}
	\ka_{\uhor} = \fr{2 - c_{14}}{(1 - c_{13})r_0}~.
	\eeq
This time, using $A_{\uhor} = 4\pi r_\uhor^2 = \pi r_0^2$ as the area of the universal horizon, we arrive at the Smarr formula for the present solution
	\beql{exact-soln:c123-0:Smarr}
	M_{\ae} = \fr{(1 - c_{13})\ka_{\uhor}A_{\uhor}}{4\pi\Gae}~.
	\eeq
Therefore, the first law of black mechanics for the present solution is
	\beql{exact-soln:c123-0:1L}
	\de M_{\ae} = \fr{(1 - c_{13})\ka_{\uhor}\,\de A_{\uhor}}{8\pi\Gae}~.
	\eeq
We find perfect agreement, once again, upon comparing \eq{exact-soln:c123-0:Smarr} and \eq{exact-soln:c123-0:1L} with the general Smarr formula \eq{Smarr} and the first law \eq{1L} respectively. This time, the interesting piece proportional to $c_{123}K_\uhor$ is absent for the obvious reason.
%***********************************************************************************************************************************************************************************************************
\section{Summary and Conclusions}
In this paper we have studied static, spherically symmetric, asymptotically flat black hole solutions of the Einstein-{\ae}ther theory, a generally covariant modification of general relativity where a vector field, the {\ae}ther, is forced to satisfy a unit normalization constraint. Since there is a preferred frame of reference defined by the {\ae}ther, the solutions of the theory violate local Lorentz invariance and so matter fields do not necessarily have a finite local limiting speed. Even though at first sight the notion of a black hole seems impossible in such a situation, earlier work \cite{EJ:aebh, BJS:aebh} provided concrete evidence in support of the existence of a single parameter family of static, spherically symmetric, asymptotically flat black hole solutions. The causal boundary in question, called the universal horizon, is a hypersurface which traps excitations travelling even at infinite local speed. 

In this work, we extend these earlier studies by demonstrating that a Smarr relation and a first law of black hole mechanics associated with the universal horizon can be found. We also provide analytical evidence for the existence of a universal horizon by constructing two exact solutions for special choices of the couplings of the Einstein-{\ae}ther theory, and verify the Smarr formula and the first law with these special cases. Critical to our proof of the first law is the fact that the solutions depend on a single parameter, namely the mass of the solution~\cite{EJ:aebh, BJS:aebh}.

The Smarr formula \eq{Smarr} and the first law \eq{1L} suggest that in the present context, the quantity $q_{\uhor}$ \eq{def:quh} plays the same role as surface gravity at the Killing horizon of a black hole in general relativity. Indeed, from \eq{def:quh} $q_{\uhor}$ does include a contribution from the surface gravity at the universal horizon, but there is also an additional piece that depends on the extrinsic curvature of the preferred foliation as it asymptotes to the universal horizon. Based on the causal nature of the universal horizon, a thermodynamic interpretation of the first law \eq{1L} seems necessary, along the lines of \cite{HawkingRad}.

Implementing such a thermodynamic interpretation in a concrete manner may, however, prove problematic. In deriving a Smarr formula and first law we have just dealt with classical, low energy physics. There is no need to worry about quantum effects in either the matter or gravitational sectors or the ultimately necessary ultraviolet completion of Einstein-{\ae}ther theory. However, if one wants to provide an explicit and consistent thermodynamic interpretation of a horizon entropy one must introduce radiation from the causal horizon and now one runs into problematic issues. In the usual picture of Hawking radiation from a black hole, finite wavelength modes at infinity originates as infinitely short wavelengths near the horizon. If exact local Lorentz invariance holds then the ultraviolet near horizon modes  are not an issue - the necessary microscopic physics is determined by the symmetry. This however, requires assumptions about Lorentz symmetry at untested energies and it is not conclusively proven that the microscopic structure of spacetime respects exact Lorentz invariance (although we have never seen a violation of Lorentz invariance~\cite{MattinglyReview, Kostelecky:2008ts}). This is the so-called ``transplanckian problem'' of black hole physics~\cite{Jacobson:1991gr}. Similarly, in the Einstein-{\ae}ther case, modes that escape to infinity have infinitely high speeds/short wavelengths with respect to the {\ae}ther frame near the universal horizon. Yet here we have neither local Lorentz symmetry nor a unique ultraviolet completion to the theory that would provide the necessary microscopic framework to unambiguously specify the near horizon mode behavior.

While there are only very few studies of the thermodynamics of universal horizons (c.f. the discussion on universal horizons in Einstein-{\ae}ther theory in~\cite{Blas:2011ni}), a number of authors have studied radiation from a Killing horizon when Lorentz symmetry is broken in the ultraviolet for the quantum field. The thermal nature of Hawking radiation from a Killing horizon has been shown to be fairly robust against ultraviolet modifications to Lorentz symmetry that yield subluminal propagation for quantum fields~\cite{Unruh:2004zk} but the behavior for superluminal fields is much less straightforward~\cite{Barcelo:2008qe}. Hence the radiation spectrum from a universal horizon, which is not a Killing horizon and where modes are necessarily ``superluminal'', is completely unknown. Finally, if Einstein-{\ae}ther is the low energy limit of a renormalizable theory such as Horava-Lifshitz gravity, then there are difficulties with assigning a holographic entropy to black holes, as this may interfere with the necessary ultraviolet behavior (c.f. the discussion in ~\cite{Shomer:2007vq}). On the other hand, simple general thermodynamic arguments imply that there should be an entropy and our first law hints that the entropy is still a function of variables on a horizon area. These are puzzles that requires further investigation, which we leave for future work.

There is one more practical complication that arises when deriving thermodynamics as in Einstein-{\ae}ther theory \v{C}erenkov radiation generically occurs since {\ae}ther-metric modes and matter fields all have different speeds. In particular, a matter field mode propagating outwards from the universal horizon would emit spin-0 \v{C}erenkov radiation, thereby modifying any thermal spectrum. Since no detailed examination of the radiation spectrum from a universal horizon has been made yet, it is unclear whether or not there is a \v{C}erenkov-type component in addition to any presumed thermal flux. Note, however, that when $c_{14}\rightarrow 0$ the speed of the spin-0 mode goes to infinity, and so there is no possibility of emission of spin-0 \v{C}erenkov radiation. On the other hand, when $c_{123} \rightarrow 0$, the speed of the spin-0 mode goes to zero, and so any spin-0 \v{C}erenkov radiation would carry no energy away from the universal horizon. Thus, in both cases  the reduction to a situation where the presumed temperature is proportional to the surface gravity is consistent with a limit where any \v{C}erenkov radiation would naturally become unimportant. 

\acknowledgments 

We are grateful to Ted Jacobson for critically reviewing an earlier draft of this paper and for providing numerous invaluable suggestions. P.B. acknowledges the hospitality of the KITPC, Beijing, the Berkeley Center for Theoretical Physics and the theory group at CERN. The work of P.B. and J.B. is supported by the NSF CAREER grant PHY-0645686. D.M. thanks the University of New Hampshire for research support.

\appendix
%***********************************************************************************************************************************************************************************************************
\section{Some algebraic details}\label{append}
In this appendix, we provide some of the technical algebraic details behind our analysis. Of central importance to our analysis is the hypersurface orthogonality relation \eq{HSO:0}, as well as the assumptions of spherical symmetry and staticity. We already noted in section \ref{sec:all-about-ae} (see the footnote around \eq{reln:K-K0-hatK}) that as a consequence of the hypersurface orthogonality of $u^a$, spherical symmetry and staticity, the vector $s^a$ is hypersurface orthogonal with respect to the foliations $\Sg_s$ of the spacetime, and satisfy $s\wg\ds = 0$. Contracting these hypersurface orthogonality conditions with $u^a$ and $s^a$ respectively, we obtain
	\beql{append:HSO:1}
	\Dl_{[a} u_{b]} = -(a\cdot s)u_{[a} s_{b]}, \qquad \Dl_{[a} s_{b]} = -K_0 u_{[a} s_{b]}
	\eeq
Further contractions of these relations with any Killing vector $\eta^a$ and with $u^a$ and/or $s^a$ lead to extremely useful scalar identities which play a major role in our analysis, and in particular, in the derivation of \eq{Smarr:EOM}. One of our goals in this appendix is to briefly outline how these identities are obtained in a generalized and coherent fashion. To that end, we recall (as already noted in the paragragh following \eq{reln:a-s}) that by the spherical symmetry of the problem, any rank-two tensor can have components only along the bi-vectors $u_a u_b$, $u_{(a} s_{b)}$, $u_{[a} s_{b]}$, $s_a s_b$ and $\hatmet_{a b}$. The most basic among these basis-expansions are those for $\Dl_a u_b$
	\beql{append:basis-expand:Dlu}
	\begin{split}
	& \Dl_a u_b = -(a\cdot s)u_a s_b + K_{a b}, \\ 
	& K_{a b} = K_0 s_a s_b + \fr{\hat{K}}{2} \hatmet_{a b}
	\end{split}
	\eeq
and for $\Dl_a s_b$
	\beql{append:basis-expand:Dls}
	\begin{split}
	& \Dl_a s_b = K_0 s_a u_b + K^{(s)}_{a b}, \\
	& K^{(s)}_{a b} = -(a\cdot s)u_a u_b + \fr{\hat{k}}{2}\hatmet_{a b}
	\end{split}
	\eeq
where $K^{(s)}_{a b}$ is the extrinsic curvature of the hypersurfaces $\Sg_s$ due to their embedding in the spacetime, and $\hat{k}$ and $\hat{K}$ are the traces of the extrinsic curvatures of the two-spheres $\Sph$ due to their embeddings in $\Sg_U$ and $\Sg_s$ respectively  
	\beql{append:def:extk-S}
	\hat{k} = (1/2)\met^{a b}\LieD_s\hatmet_{a b}, \qquad \hat{K} = (1/2)\met^{a b}\LieD_u\hatmet_{a b}
	\eeq
Our results in this paper make heavy uses of \eq{append:basis-expand:Dlu} and \eq{append:basis-expand:Dls}.

Now, consider an arbitrary vector of the following form
	\beql{append:def:Aa}
	A_a = -f u_a + h s_a
	\eeq
where $f$ and $h$ are arbitrary functions respecting the symmetries of the problem. A natural construct is the two-form
	\beql{append:def:Fab}
	F_{a b} = \Dl_{[a} A_{b]} = Q\,u_{[a} s_{b]}
	\eeq
where the second equality follows by the spherical symmetry of the problem, and the scalar $Q$ is given by
	\beql{append:def:Q}
	Q = f(a\cdot s) + \Dl_s f - hK_0 - \Dl_u h
	\eeq
Using the torsion-free condition of the covariant derivative, it can be shown that $Q$, like $F_{a b}$, is invariant under the \tql{gauge transformations} $A_a \mapsto A'_a = A_a + \Dl_a\La$. 

Comparing \eq{append:HSO:1} with \eq{append:def:Fab}, we see that the former relations are special cases of the latter. Contracting \eq{append:def:Fab} with a Killing vector $\eta^a$ yields
	\beql{append:genHSO:2}
	\Dl_a(A\cdot\eta) = Q(s\cdot\eta)u_a - Q(u\cdot\eta)s_a
	\eeq
Contracting \eq{append:genHSO:2} further with $u^a$ and $s^a$ gives
	\beql{append:genHSO:3}
	\Dl_u (A\cdot\eta) = -Q(s\cdot\eta), \qquad \Dl_s (A\cdot\eta) = -Q(u\cdot\eta)
	\eeq
We make ample use of the special cases of \eq{append:genHSO:2} and \eq{append:genHSO:3} for $A^a = u^a$ and $A^a = s^a$ throughout the analysis.

The flux of $F_{a b}$ over any two-sphere $\Sph$ is
	\beql{append:Fab:flux}
	\int\limits_{\Sph}\dSg^{a b}F_{a b} = \int\limits_{\Sph}\dA\, Q
	\eeq
Since $F_{a b}$ is antisymmetric, we have a {\it kinematically} conserved current $J^a$ through
	\beql{append:def:ja-Fab}
	\Dl_b F^{a b} = J^a
	\eeq
We call $J^a$ a kinematically conserved current since it satisfies the continuity equation $\Dl\cdot J = 0$ irrespective of any equations of motion.

By spherical symmetry, the current can only have components along $u^a$ and $s^a$. These projections are given by
	\beql{append:ju-js}
	\begin{split}
	& (u\cdot J) = -(\hat{k}Q + \Dl_s Q) = -\vecDl\cdot[Q\vec{s}]~, \\
	& (s\cdot J) = -(\hat{K}Q + \Dl_u Q) = -\uvecDl\cdot[Q\uvec{u}]
	\end{split}
	\eeq
where $\vec{v}^a = \tn{\proj}{^a_b}v^b$ and $\uvec{v}^a = \tn{\sproj}{^a_b}v^b$ are projections of any four vector $v^a$ on the hypersurfaces $\Sg_U$ and $\Sg_s$ respectively, $\tn{\proj}{^a_b} = u^a u_b + \tn{\de}{^a_b}$ and $\tn{\sproj}{^a_b} = -s^a s_b + \tn{\de}{^a_b}$ are the corresponding projection tensors, and $\vecDl_a = \tn{\proj}{_a^b}\Dl_b$ and $\uvecDl_a = \tn{\sproj}{_a^b}\Dl_b$ are the corresponding projections of the covariant derivatives.

The most important example of a kinematically conserved current, which is central in the derivation of \eq{Smarr:EOM}, is that due to the timelike Killing vector $\chi^a$. Choosing the vector $A_a = (1/2)\chi_a$ in \eq{append:def:Aa}, so that from \eq{append:def:Fab} $F_{a b} = \Dl_a\chi_b$, we obtain $Q = -\ka$. From the identity $\Dl_b \Dl^a \chi^b = \tn{\Ric}{^a_b}\chi^b$, which forms the basis of the Smarr relation, we find that the kinematically conserved current is $J_\chi^a = \tn{\Ric}{^a_b}\chi^b$ such that
	\beql{append:def:Jchi}
	(u\cdot J_\chi) = \vecDl\cdot[\ka\vec{s}], \qquad (s\cdot J_\chi) = \uvecDl\cdot[\ka\uvec{u}]
	\eeq
We conclude this appendix by noting the following two identities 
	\beql{append:(u.X)Dlv:Sg}
	(u\cdot\eta)(\Dl\cdot v) = \vecDl\cdot\lf[\Big\{-(u\cdot v)(s\cdot\eta) + (s\cdot v)(u\cdot\eta)\Big\}\vec{s}\rf]
	\eeq
and
	\beql{append:(s.X)Dlv:Sgs}
	(s\cdot\eta)(\Dl\cdot v) = \uvecDl\cdot\lf[\Big\{-(u\cdot v)(s\cdot\eta) + (s\cdot v)(u\cdot\eta)\Big\}\uvec{u}\rf]
	\eeq
valid for any four-vector $v^a$. The identities \eq{append:(u.X)Dlv:Sg} and \eq{append:(s.X)Dlv:Sgs} are indispensible for dealing with the contributions from the {\ae}ther in the derivation of \eq{Smarr:EOM}.
%***********************************************************************************************************************************************************************************************************
\section{More on spin-0 horizon (ir)regularity}\label{append:spin-0}
In this appendix, we present some further studies of the equations of motion, where the irregularity of a general {\ae}ther black hole solution near its spin-0 horizon for generic values of $r_0$ and $r_{\ae}$~\cite{EJ:aebh, BJS:aebh} manifests itself in interesting ways.

We have performed a near-universal-horizon analysis of the equations of motion to  lowest order in the near-universal-horizon radial coordinate. The most striking feature of the analysis is that two separate treatments are required for the cases $c_{14} \neq 0$ and $c_{14} = 0$, since, the series coefficients are divergent in $c_{14}$ in the $c_{14} \neq 0$ case\ft{The situation should be compared with the asymptotic analysis when $c_{123} \neq 0$ and the corresponding divergence of the asymptotic solution in $c_{123}$ -- see the comments following equation \eq{asymptotic-behaviour:aether}. As discussed earlier, for small $c_{123}$ the spin-0 horizon is pushed towards asymptotic infinity. Therefore the asymptotic solution is a good approximation to the near-spin-0-horizon region of the solution for sufficiently small values of $c_{123}$. The divergence here is therefore a manifestation of the irregularity of the general asymptotic solution (where no regularity is imposed) at its spin-0 horizon.}. Indeed, in this case the near-univeral-horizon region is also the near-spin-0-horizon region when $c_{14}$ is sufficiently small. Since we did not impose any {\it a priori} regularity condition on the solution in our analysis, this divergent behaviour is clearly due to the irregularity of the solution at the spin-0 horizon.

The near-universal-horizon analysis for $c_{14} = 0$, on the other hand, successfully reproduces\ft{Of course, the exact values of the functions at the universal horizon, as well as the radial location of the universal horizon in the Eddington-Finklestein radial coordinate \eq{met:static-sph-sym}, depend on the normalization of the asymptotic data and the actual definition of the radial coordinate itself, and cannot be \tql{derived} in this kind of an analysis.} the near-universal-horizon behaviour of the exact solution for $c_{14} = 0$. Likewise, upon taking the limit $c_{123} \to 0$ of the near-universal-horizon solution for $c_{14} \neq 0$, we successfully reproduce the near-universal-horizon behaviour of the $c_{123} = 0$ exact solution.

We have also studied perturbative corrections to the exact solutions for $c_{14} = 0$ \eq{exact-soln:c14-0:e-f-u} and $c_{123} = 0$ \eq{exact-soln:c123-0:e-f-u} -- for small $c_{14}$ in the former case, and for small $c_{123}$ in the latter. The former analysis (i.e around the $c_{14} = 0$ exact solution) did not yield a complete solution because of the very complicated nature of the relevant equations -- we could only find the $\ord(c_{14})$ correction to the function $f(r)$. In the context of the $c_{123} = 0$ exact solution however, with an additional mild assumption, namely $c_{14} = 2c_{13} - \ep$ where $c_{123} = \ep \ll 1$, we found the $\ord(c_{123})$ correction to all the three basic functions $e(r)$, $f(r)$ and $\al(r)$\ft{In our analysis, the spin-0 speed $s_0$ can be kept arbitrarily small as long as $c_{123} \ll c_{13} < 1$; therefore, the spin-0 horizon can be \tql{arbitrarily far away}, i.e., \tql{arbitrarily close} to the asymptotic boundary. Looking at the expression for $r_u$ \eq{def:ru} we further notice that the assumption on $c_{14}$ implies that the perturbative correction we are studying can also be thought of as a perturbation around a Schwarzschild background. We also refer the reader to a useful study of non-spherical perturbations about black hole solutions in extended Ho\v{r}ava gravity and Einstein-{\ae}ther theory~\cite{Blas:2011ni}.}. The most important property of the perturbative corrections, relevant for the present discussion, is that the corrections have logarithmic singularities at the location of the {\it unperturbed} spin-0 horizon\ft{Even though our analysis of the small $c_{14}$ perturbation around the $c_{14} = 0$ exact solution is incomplete, the statement is true for the $\ord(c_{14})$ correction for the function $f(r)$ -- see below. The relevant equations in this context also indicate that such logarithmic singularities will be present for $e(r)$ and $\al(r)$.}. More specifically, the $\ord(c_{123})$ correction functions to the $c_{123} = 0$ exact solution depend on $\log(r/r_0)$\ft{One might be worried that the $\ord(c_{123})$ correction functions to the $c_{123} = 0$ solution destroy the asymptotic flatness condition, due to the presence of the $\log(r/r_0)$ in their expressions. We however do not claim that the $\ord(c_{123})$ correction approximates the actual solution {\it everywhere} in spacetime. Indeed, the $\ord(c_{123})$ correction can be a good approximation near the universal and the Killing horizons, but we need to consider higher order corrections in $c_{123}$ to approximate the actual solution as we move out in $r$, and presumably we need to go to all orders in $c_{123}$ in order to \tql{approximate} the actual solution at asymptotic infinity. On the other hand, by the results of~\cite{EJ:aebh, BJS:aebh} we are guaranteed to reach an asymptotically flat regular solution (provided we choose the appropriate  boundary conditions for the perturbation at the universal horizon).}, while for the $c_{14} = 0$ exact solution, the $\ord(c_{14})$ correction to $f(r)$ depends on $\log(1 - 3r_0/4r)$. Now, although the exact solutions are regular everywhere, their first order corrected counterparts are {\it a priori} not so. The logarithmic divergence at the unperturbed spin-0 horizon therefore clearly signals an irregular behaviour\ft{The irregularity occurs at the unperturbed spin-0 horizon instead of at the true spin-0 horizon of the actual solution, because, the first order corrections that we have considered here do not approximate the solution near the spin-0 horizon for the actual solution (see the previous footnote), and one needs to consider successively higher order corrections to the exact solutions to see the irregularity at the correct location properly.}.
%**********************************************************************************************************************************************************************************************************


\begin{thebibliography}{99}
\bibitem{BCH} J.~M.~Bardeen, B.~Carter, S.~W.~Hawking, ``The Four laws of black hole mechanics,'' Commun.\ Math.\ Phys.\  {\bf 31}, 161-170 (1973).
\bibitem{HawkingRad} S.~W.~Hawking, ``Particle Creation by Black Holes,'' Commun.\ Math.\ Phys.\  {\bf 43}, 199 (1975) [Erratum-ibid.\  {\bf 46}, 206 (1976)].
\bibitem{Bekenstein:1973ur} J.~D.~Bekenstein, ``Black holes and entropy,'' Phys.\ Rev.\ D {\bf 7}, 2333 (1973).
\bibitem{'tHooft} G.~'t Hooft, ``Dimensional reduction in quantum gravity,'' gr-qc/9310026.
\bibitem{Susskind} L.~Susskind, ``The World as a hologram,'' J.\ Math.\ Phys.\  {\bf 36}, 6377 (1995) [hep-th/9409089].
\bibitem{Maldacena} J.~M.~Maldacena, ``The Large N limit of superconformal field theories and supergravity,'' Adv.\ Theor.\ Math.\ Phys.\  {\bf 2}, 231 (1998) [Int.\ J.\ Theor.\ Phys.\  {\bf 38}, 1113 (1999)] [hep-th/9711200].
\bibitem{Gubser} S.~S.~Gubser, I.~R.~Klebanov and A.~M.~Polyakov, ``Gauge theory correlators from noncritical string theory,'' Phys.\ Lett.\ B {\bf 428}, 105 (1998) [hep-th/9802109].
\bibitem{Witten} E.~Witten, ``Anti-de Sitter space and holography,'' Adv.\ Theor.\ Math.\ Phys.\  {\bf 2}, 253 (1998) [hep-th/9802150].
\bibitem{J:EEqstate} T.~Jacobson, ``Thermodynamics of space-time: The Einstein equation of state,'' Phys.\ Rev.\ Lett.\  {\bf 75}, 1260 (1995) [gr-qc/9504004].
\bibitem{Verlinde} E.~P.~Verlinde, ``On the Origin of Gravity and the Laws of Newton,'' JHEP {\bf 1104}, 029 (2011) [arXiv:1001.0785 [hep-th]].
\bibitem{Padmanabhan} T.~Padmanabhan, ``Gravity and the thermodynamics of horizons,'' Phys.\ Rept.\  {\bf 406}, 49 (2005) [gr-qc/0311036].
\bibitem{Jacobson:1993vj} T.~Jacobson, G.~Kang and R.~C.~Myers, ``On black hole entropy,'' Phys.\ Rev.\ D {\bf 49}, 6587 (1994) [gr-qc/9312023].
\bibitem{JM:ae:intro} T.~Jacobson and D.~Mattingly, ``Gravity with a dynamical preferred frame,'' Phys.\ Rev.\ D {\bf 64} (2001) 024028 [gr-qc/0007031].
\bibitem{Jacobson:2000gw} T.~Jacobson and D.~Mattingly, ``Generally covariant model of a scalar field with high frequency dispersion and the cosmological horizon problem,'' Phys.\ Rev.\ D {\bf 63}, 041502 (2001) [hep-th/0009052].
\bibitem{JM:aewaves} T.~Jacobson and D.~Mattingly, ``Einstein-{\ae}ther waves,'' Phys.\ Rev.\ D {\bf 70}, 024003 (2004) [gr-qc/0402005].
\bibitem{J:ae:constraints} T.~Jacobson, ``Einstein-{\ae}ther gravity: Theory and observational constraints,'' arXiv:0711.3822 [gr-qc].
\bibitem{J:ae:status} T.~Jacobson, ``Einstein-{\ae}ther gravity: A Status report,'' PoS QG {\bf -PH} (2007) 020 [arXiv:0801.1547 [gr-qc]].
\bibitem{GJ:+E} D.~Garfinkle, T.~Jacobson, ``A positive energy theorem for Einstein-{\ae}ther and Ho\v{r}ava gravity,'' arXiv:1108.1835 [gr-qc]].
\bibitem{Eling:2004dk} C.~Eling, T.~Jacobson and D.~Mattingly, ``Einstein-{\ae}ther theory,'' gr-qc/0410001.
\bibitem{MattinglyReview} D.~Mattingly, ``Modern tests of Lorentz invariance,'' Living Rev.\ Rel.\  {\bf 8}, 5 (2005) [gr-qc/0502097].
\bibitem{EJ:aebh} C.~Eling, T.~Jacobson, ``Black Holes in Einstein-{\ae}ther Theory,'' Class.\ Quant.\ Grav.\  {\bf 23}, 5643-5660 (2006), [gr-qc/0604088].
\bibitem{BJS:aebh} E.~Barausse, T.~Jacobson, T.~P.~Sotiriou, ``Black holes in Einstein-{\ae}ther and Ho\v{r}ava-Lifshitz gravity,'' Phys.\ Rev.\  {\bf D83}, 124043 (2011), [arXiv:1104.2889 [gr-qc]].
\bibitem{Dubovsky:2006vk} S.~L.~Dubovsky and S.~M.~Sibiryakov, ``Spontaneous breaking of Lorentz invariance, black holes and perpetuum mobile of the 2nd kind,'' Phys.\ Lett.\ B {\bf 638}, 509 (2006) [hep-th/0603158].
\bibitem{Eling:2007qd} C.~Eling, B.~Z.~Foster, T.~Jacobson and A.~C.~Wall, ``Lorentz violation and perpetual motion,'' Phys.\ Rev.\ D {\bf 75}, 101502 (2007) [hep-th/0702124 [HEP-TH]].
\bibitem{Jacobson:2008yc} T.~Jacobson and A.~C.~Wall, ``Black Hole Thermodynamics and Lorentz Symmetry,'' Found.\ Phys.\  {\bf 40}, 1076 (2010) [arXiv:0804.2720 [hep-th]].
\bibitem{Carroll:2004ai} S.~M.~Carroll and E.~A.~Lim, ``Lorentz-violating vector fields slow the universe down,'' Phys.\ Rev.\ D {\bf 70}, 123525 (2004) [hep-th/0407149].
\bibitem{Elliott:2005va} J.~W.~Elliott, G.~D.~Moore and H.~Stoica, ``Constraining the new Aether: Gravitational Cerenkov radiation,'' JHEP {\bf 0508}, 066 (2005) [hep-ph/0505211].
\bibitem{Horava} P.~Ho\v{r}ava, ``Quantum Gravity at a Lifshitz Point,'' Phys.\ Rev.\ D {\bf 79} (2009) 084008 [arXiv:0901.3775 [hep-th]].
\bibitem{J:hso-ae=hor} T.~Jacobson, ``Extended Ho\v{r}ava gravity and Einstein-{\ae}ther theory,'' Phys.\ Rev.\ D {\bf 81}, 101502 (2010) [Erratum-ibid.\ D {\bf 82}, 129901 (2010)] [arXiv:1001.4823 [hep-th]].
\bibitem{Jacobson-Parentani} T.~Jacobson and R.~Parentani, ``Horizon surface gravity as 2d geodesic expansion,'' Class.\ Quant.\ Grav.\  {\bf 25}, 195009 (2008) [arXiv:0806.1677 [gr-qc]].
\bibitem{J:Ceqn} T.~Jacobson, ``Initial value constraints with tensor matter,'' [arXiv:1108.1496 [gr-qc]].
\bibitem{Eling:aeE} C.~Eling, ``Energy in the Einstein-{\ae}ther theory,'' Phys.\ Rev.\ D {\bf 73}, 084026 (2006) [Erratum-ibid.\ D {\bf 80}, 129905 (2009)] [gr-qc/0507059].
\bibitem{Foster:aeNQ} B.~Z.~Foster, ``Noether charges and black hole mechanics in Einstein-{\ae}ther theory,'' Phys.\ Rev.\ D {\bf 73}, 024005 (2006) [gr-qc/0509121].
\bibitem{Townsend} See any standard discussion on the derivation of the laws of black hole mechanics via the scaling argument, e.g., P.~K.~Townsend, ``Black holes: Lecture notes,'' gr-qc/9707012, page 113.
\bibitem{Wald:NQ} R.~M.~Wald, ``Black hole entropy is the Noether charge,'' Phys.\ Rev.\ D {\bf 48}, 3427 (1993) [gr-qc/9307038].
\bibitem{Foster:field-redef} B.~Z.~Foster, ``Metric redefinitions in Einstein-{\ae}ther theory,'' Phys.\ Rev.\ D {\bf 72}, 044017 (2005) [gr-qc/0502066].
\bibitem{Kostelecky:2008ts} V.~A.~Kostelecky and N.~Russell, ``Data Tables for Lorentz and CPT Violation,'' Rev.\ Mod.\ Phys.\  {\bf 83}, 11 (2011) [arXiv:0801.0287 [hep-ph]].
\bibitem{Jacobson:1991gr} T.~Jacobson, ``Black hole evaporation and ultrashort distances,'' Phys.\ Rev.\ D {\bf 44}, 1731 (1991).
\bibitem{Blas:2011ni} D.~Blas and S.~Sibiryakov, ``Ho\v{r}ava gravity versus thermodynamics: The Black hole case,'' Phys.\ Rev.\ D {\bf 84}, 124043 (2011) [arXiv:1110.2195 [hep-th]].
\bibitem{Unruh:2004zk} W.~G.~Unruh and R.~Schutzhold, ``On the universality of the Hawking effect,'' Phys.\ Rev.\ D {\bf 71}, 024028 (2005) [gr-qc/0408009].
\bibitem{Barcelo:2008qe} C.~Barcelo, L.~J.~Garay and G.~Jannes, ``Sensitivity of Hawking radiation to superluminal dispersion relations,'' Phys.\ Rev.\ D {\bf 79}, 024016 (2009) [arXiv:0807.4147 [gr-qc]].
\bibitem{Shomer:2007vq} A.~Shomer, ``A Pedagogical explanation for the non-renormalizability of gravity,'' arXiv:0709.3555 [hep-th].
\end{thebibliography}
\end{document}